\documentclass[12pt,twoside]{article} 
\usepackage[latin1]{inputenc}
\usepackage{graphicx}
\usepackage{enumitem}
\usepackage{latexsym}
\usepackage{amssymb}
\usepackage{epsfig}
\usepackage{xcolor}
\usepackage{amsmath}
\usepackage{float}

\allowdisplaybreaks 

\numberwithin{equation}{section}

\def\eps{\varepsilon}
\def \R{\mathbb{R}}
\def \N{\mathbb{N}}

\newcommand{\ns}{\lfloor ns\rfloor}

\newcommand{\ceil}[1]{\lceil #1 \rceil}
\newcommand{\floor}[1]{\lfloor #1 \rfloor}

\def\en{\mathbb{N}}

\def\er{\mathbb{R}}

\def\cf{\mathcal{F}}
\def\e{\varepsilon}
\def\he{\hat{\varepsilon}}
\def\ben{\bar{\epsilon}_n}

\def\sjn{\sum_{j=1}^n}
\def\op{o_P\Big(\frac 1{\sqrt n}\Big)}

\def\beq{\begin{eqnarray*}}
\def\eeq{\end{eqnarray*}}
\def\farc{\frac}
\def\sjns{\sum_{j=1}^{\lfloor ns\rfloor}}

\def\ns{\lfloor ns\rfloor}

\def\nn{\nonumber}
\newcommand{\nto}{\xrightarrow[n\to\infty]{}}

\newcommand{\gts}{{\tilde g_n^*}}
\newcommand{\thet}{{\vartheta}}

\begin{document}

\title{\bf Hypotheses tests in boundary regression models
}

\author{\sc Holger Drees, Natalie Neumeyer and Leonie Selk$^*$\\
University of Hamburg, Department of Mathematics \\
Bundesstrasse 55, 20146 Hamburg, Germany\\
}

\maketitle

\newtheorem{theo}{Theorem}[section]
\newtheorem{lemma}[theo]{Lemma}
\newtheorem{cor}[theo]{Corollary}
\newtheorem{rem}[theo]{Remark}
\newtheorem{prop}[theo]{Proposition}
\newtheorem{defin}[theo]{Definition}
\newtheorem{example}[theo]{Example}

\begin{abstract}
Consider a nonparametric regression model with one-sided errors and regression function in a general H\"{o}lder class.
We estimate the regression function via minimization of the local
integral of a polynomial approximation. We show uniform rates of
convergence for the simple  regression estimator as well as for a
smooth version. These rates carry over to mean regression models
with a symmetric and bounded error distribution. In such a setting,
one obtains faster rates for irregular error distributions
concentrating sufficient mass near the endpoints than for the usual
regular distributions. The results are applied to prove asymptotic
$\sqrt{n}$-equivalence of a residual-based (sequential) empirical
distribution function to the (sequential) empirical distribution
function of unobserved errors in the case of irregular error
distributions. This result is remarkably different from
corresponding results in mean regression with regular errors. It can
readily be applied to develop goodness-of-fit tests for the error
distribution. We present some examples and investigate the small
sample performance in a simulation study. We further discuss
asymptotically distribution-free hypotheses tests for independence
of the error distribution from the points of measurement and for
monotonicity of the boundary function as well.
\end{abstract}




\section{Introduction}

We consider boundary regression models
of the form
$$ Y_i = g(x_i)+\eps_i, \quad i=1,\dots,n,$$
with negative errors $\eps_i$ whose survival function $1-F(y)$
behaves like a multiple of $|y|^\alpha$ for some $\alpha>0$ near the
origin. Such models naturally arise in image analysis, analysis of
auctions and records, or in extreme value analysis with covariates.
 For such a boundary
regression model with multivariate random covariates and twice
differentiable regression function, Hall and Van Keilegom (2009)
establish a minimax rate for estimation of $g(x)$ (for fixed $x$)
under quadratic loss and determine pointwise asymptotic
distributions of an estimator which is defined as a solution of a
linear optimization problem (cf.\ Remark \ref{rem:HallKeilegom}).
Relatedly,
 M\"{u}ller
and Wefelmeyer (2010) consider a mean regression model with
(unknown) symmetric  support of the error distribution and H\"{o}lder
continuous regression function. They discuss pointwise MSE rates for
estimators of the regression function that are defined as the
average of local maxima and local minima. Meister and Rei{\ss} (2013)
consider a regression model with known bounded support of the
errors. They show asymptotic equivalence in the strong LeCam sense
to a continuous-time Poisson point process model when the error
density has a jump at the endpoint of its support. For a regression
model with error distribution that is one-sided and regularly
varying at 0 with index $\alpha>0$, Jirak et al.\ (2014)  suggest an
estimator for the boundary regression function which adapts
simultaneously to the unknown smoothness of the regression function
and to the unknown extreme value index $\alpha$. Rei{\ss} and Selk
(2016+) construct efficient and unbiased estimators of linear
functionals of the regression function in the case of exponentially
distributed errors as well as in the limiting  Poisson point process
experiment by Meister and Rei{\ss} (2013).

Closely related to regression estimation in models with one-sided
errors is the estimation of a boundary function $g$ based on a
sample from $(X,Y)$ with support $\{(x,y)\in [0,1]\times
[0,\infty]\mid y\leq g(x)\}$. For such models, H\"{a}rdle et al.\ (1995)
and Hall et al.\ (1998) proved minimax rates both for $g(x)$ and for
the $L_1$-distance between $g$ and its estimator. Moreover, they
showed that an approach using local polynomial approximations of $g$
yields this optimal rate.  Explicit estimators in terms of higher
order moments were proposed and analyzed by Girard and Jacob (2008)
and Girard et al.\ (2013). Daouia et al.\ (2016) consider spline
estimation of a support frontier curve and obtain uniform rates of
convergence.

The aim of the paper is to develop tests for model assumptions  in
boundary regression models. In particular we will suggest
asymptotically distribution-free tests for
\begin{itemize}
\item parametric classes of error distributions (goodness-of-fit)
\item independence of the error distribution from the points of measurement
\item monotonicity of the boundary function.
\end{itemize}
The test statistics are based on (sequential) empirical processes of
residuals. To investigate these, we need uniform rates of
convergence for the regression estimator, which are of interest on
its own. To our knowledge, uniform rates so far have only been shown
by Daouia et al.\ (2016) who do not obtain optimal rates. Our
results can also be applied to mean regression models with bounded
symmetric error distribution. For regression functions $g$ in a
H\"{o}lder class of order $\beta$, we obtain the rate $((\log
n)/n)^{\beta/(\alpha\beta+1)}$. Thus, for tail index $\alpha\in
(0,2)$ of the error distribution, the rate is  faster than the
typical rate one has in mean regression models with regular errors.
For pointwise and $L^p$-rates of convergence, it has been known in
the literature that faster rates are possible for nonparametric
regression estimation in models with irregular error distribution,
see e.g.\ Gijbels and Peng (2000), Hall and Van Keilegom (2009), or
M\"{u}ller and Wefelmeyer (2010).

The uniform rate of convergence  for the regression estimator
enables us to derive asymptotic expansions for residual-based
empirical distribution functions and to prove weak convergence of
the residual-based (sequential) empirical distribution function. We
state conditions under which the influence of the regression
estimation is negligible such that the same results are obtained as
in the  case of observable errors.  We apply the results to
derive goodness-of-fit tests for parametric classes of error
distributions. Asymptotic properties of residual empirical
distribution functions in mean regression models were investigated
by Akritas and Van Keilegom (2001), among others. As  the regression
estimation strongly influences the asymptotic behavior of the
empirical distribution function in these regular models, asymptotic
distributions of goodness-of-fit test statistics are involved, and
typically bootstrap is applied to obtain critical values, see
Neumeyer et al.\ (2006). In contrast, in the present situation with
an irregular error distribution, standard critical values can be
used.

In nonparametric frontier models,  Wilson (2003)  discusses several
possible tests for assumptions of independence, for instance
independence between input levels  and output inefficiency. Those
assumptions are needed to prove validity of  bootstrap procedures
and are thus crucial in applications, but they may be violated; see
Simar and Wilson (1998). Wilson (2003) points out the analogy to
tests for independence between errors and covariates in regression
models, but no asymptotic distributions  are derived. Tests for
independence in nonparametric mean and quantile regression models
that are similar to the test we will consider are suggested by
Einmahl and Van Keilegom (2008) and Birke et al.\ (2016+).

There is an extensive literature on regression with one-sided  error
distributions and similar models (in particular production frontier
models) which assume monotonicity of the boundary function, see
Gijbels et al.\ (1999), the literature cited therein  and the
monotone nonparametric maximum likelihood estimator in Rei{\ss} and Selk
(2016+). Monotonicity of a production frontier function in each
component is given under the strong disposability assumption, but
may often not be fulfilled; see e.g.\ F\"{a}re and Grosskopf (1983). We
are not aware of hypothesis tests for monotonicity or other shape
constraints in the context of boundary regression, but would like to
mention Gijbels' (2005) review on testing for monotonicity in mean
regression. Tests similar in spirit to the one we are suggesting
here were considered by Birke and Neumeyer (2013) and  Birke et al.\
(2016+) for mean and quantile regression models, respectively.

The remainder of the article is organized as follows. In Section 2
the regression model under consideration is presented  and model
assumptions are formulated. The regression estimator is defined  and
uniform rates of convergence are given. A smooth modification of the
estimator is considered and uniform rates of convergence for this
estimator as well as its derivative are shown. In Section 3 residual
based empirical distribution functions based on both regression
estimators are investigated. Conditions are stated under which the
influence of regression estimation is asymptotically
$\sqrt{n}$-negligible. Furthermore, an expansion of the residual
empirical distribution function is shown that is valid under more
general conditions. In Section 4 goodness-of-fit tests for the error
distribution are discussed in general and in some detailed examples.
We investigate  the finite sample performance of the tests in a
small simulation study. We further discuss hypotheses tests for
independence of the error distribution from the design points as
well as a test for monotonicity of the boundary function. All proofs
are given in the appendix.

\section{The regression function: uniform rates of convergence}

We consider a regression model with fixed equidistant design and one-sided errors,
\begin{eqnarray}\label{model}
Y_i&=&g(\textstyle{\frac{i}{n}})+\eps_i,\quad i=1,\dots,n,
\end{eqnarray}
under the following assumptions:

\begin{enumerate}[label=(\textbf{F\arabic{*}})]
\item \label{F1} The errors $\eps_1,\dots,\eps_n$ are independent and
identically distributed and supported on $(-\infty,0]$. The error distribution function fulfills
$$F(y)=1-c|y|^\alpha+r(y), \quad y<0,$$
for some $\alpha>0$, with $r(y)=o(|y|^\alpha)$ for $y\nearrow 0$.
\end{enumerate}
\begin{enumerate}[label=(\textbf{G\arabic{*}})]
\item \label{G1} The regression function $g$ belongs to some H\"{o}lder class of order $\beta\in (0,\infty)$,
 i.\,e.\ $g$ is $\lfloor \beta\rfloor$-times differentiable on $[0,1]$ and the $\lfloor \beta\rfloor$-th derivative satisfies
$$c_g:= \sup_{t,x\in[0,1]\atop t\neq x}\frac{|g^{(\lfloor \beta\rfloor)}(t)-g^{(\lfloor \beta\rfloor)}(x)|}{|t-x|^{\beta-\lfloor \beta\rfloor}}<\infty.$$
\end{enumerate}

\begin{figure}[h]

\hspace*{1cm}$\alpha=0.5$\hspace{3cm}$\alpha=1$\hspace{3cm}$\alpha=2$\hspace{3cm}$\alpha=3$
\\
\epsfig{file=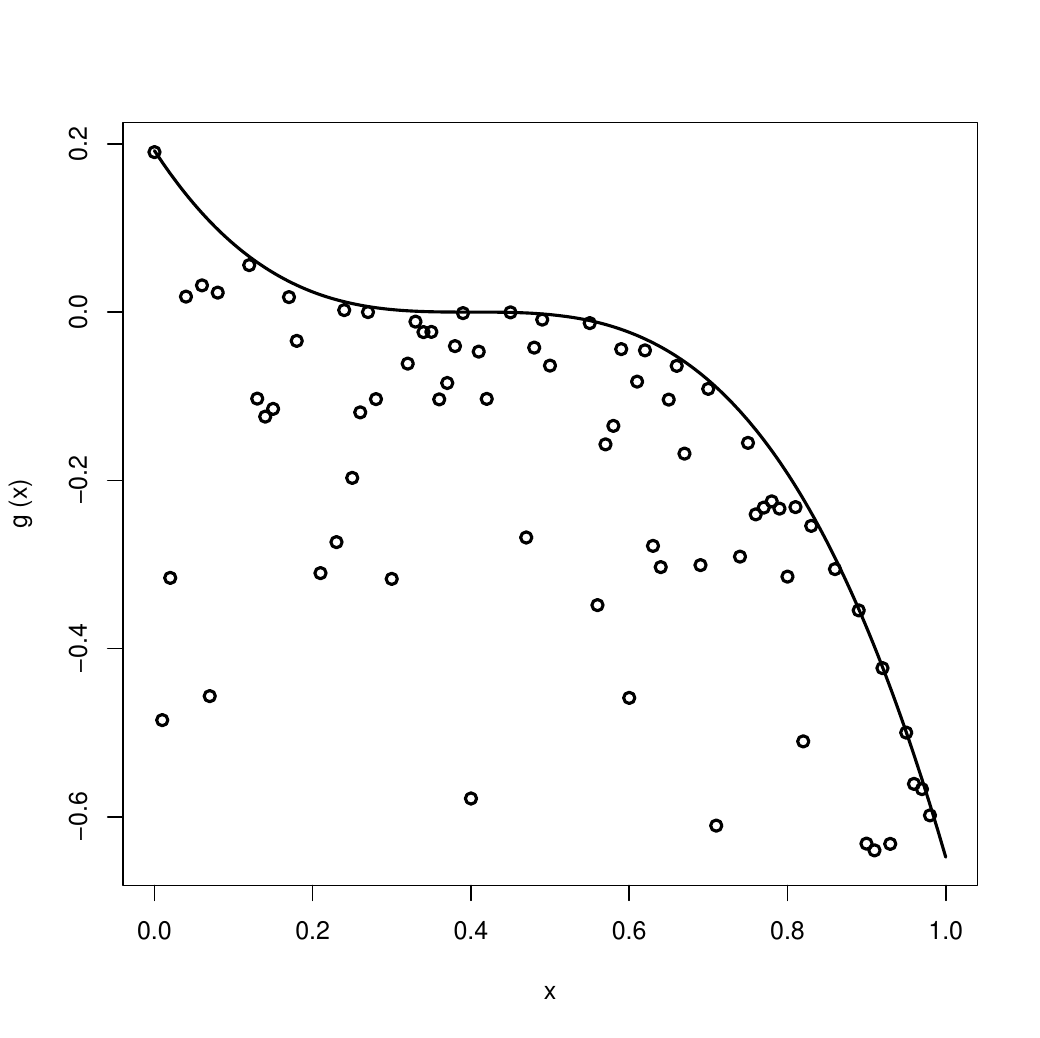,width=4cm}\epsfig{file=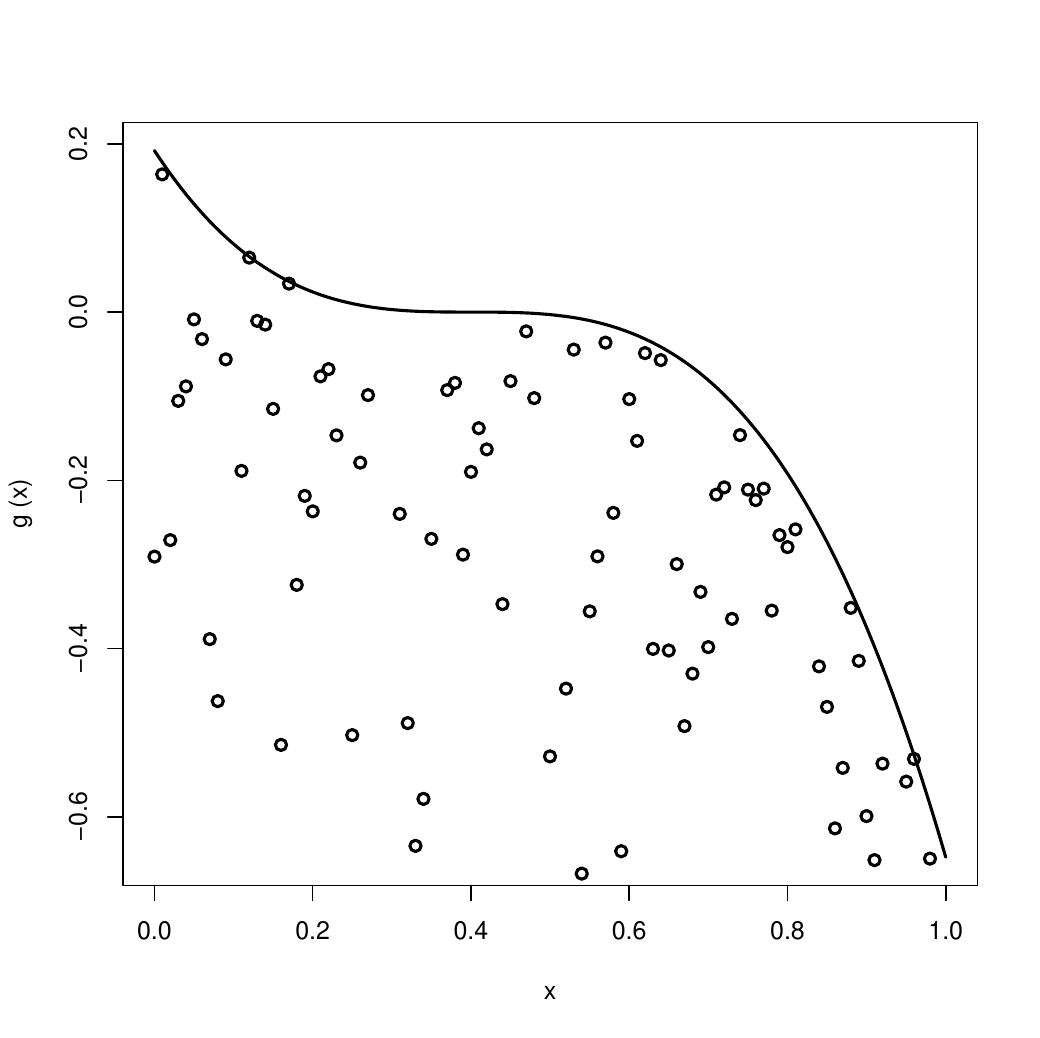,width=4cm}\epsfig{file=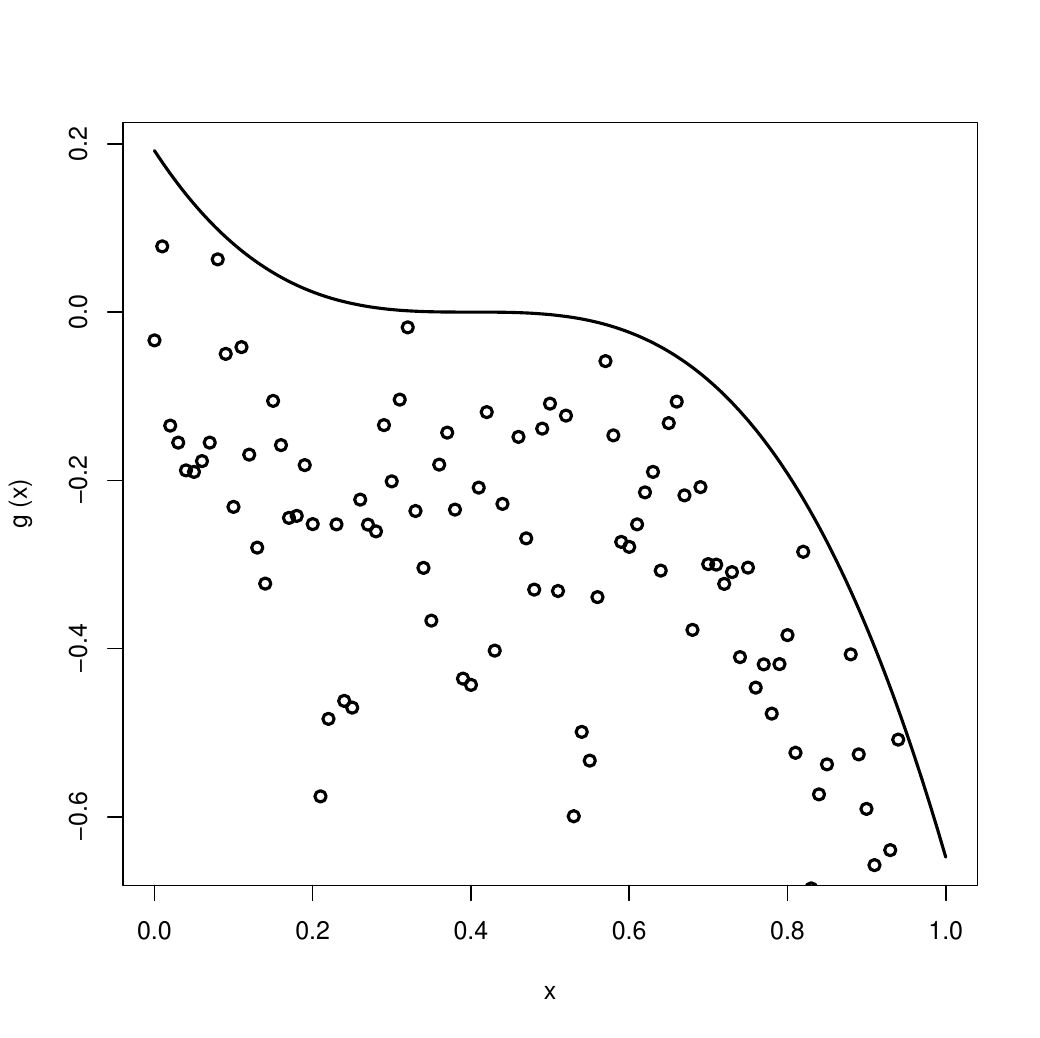,width=4cm}
\epsfig{file=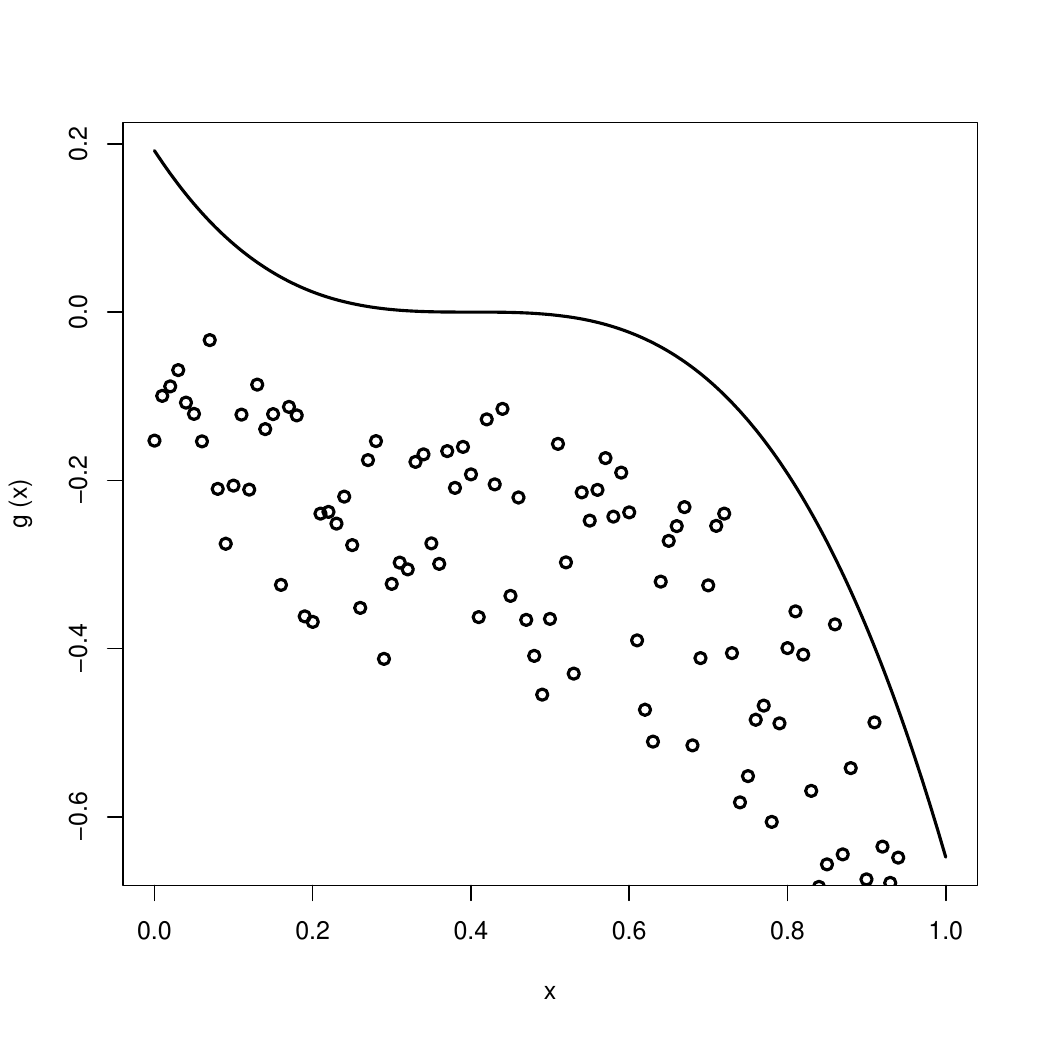,width=4cm}
\caption{\label{fig-grafiken}\it Scatter plots of $(\textstyle{\frac{i}{n}},Y_i)$, $i=1,\dots,n$, and the true
 regression function $g(x)=-3(x-0.4)^3$. The error distribution is Weibull
 $F(y)=\exp(-(|y|/\theta)^\alpha)I_{(-\infty,0)}(y)+I_{[0,\infty)}(y)$ with scale $\theta=0.3$ and shape parameter $\alpha$.}

\end{figure}

In Figure \ref{fig-grafiken} some scatter plots of data according to
model \eqref{model} are shown for different tail indices $\alpha$ of
the error distribution.

\begin{rem}\rm
 Strictly speaking, we consider a triangular scheme in \eqref{model}, and the errors $\eps_i$
depend on $n$ too, as the $i$th regression point $i/n$ varies with
$n$. For notational simplicity, we suppress the second index,
because the distribution of the errors does not depend on $n$. $\blacksquare$
\end{rem}

We consider an estimator that locally approximates the regression
function by a polynomial  while lying above the data points. More
specifically, for $x\in[0,1]$, let
$$ \hat g_n(x) := \hat g(x) := p(x) $$
 where $p$ is a polynomial of order $\lceil\beta\rceil -1$ and minimizes the local integral
\begin{equation}\label{hat-g}
\int_{x-h_n}^{x+h_n} p(t)\,dt
\end{equation}
under the constraints $p(\frac jn)\geq Y_j$ for all $j\in\{1,\dots,n\}$ such that $|\frac jn-x|\leq h_n$.
For the asymptotic analysis of this estimator, we  need the following assumption:
\begin{enumerate}[label=(\textbf{H\arabic{*}})]
\item \label{H1} Let $(h_n)_{n\in\mathbb{N}}$ be a sequence of positive bandwidths that satisfies
 $\lim_{n\to\infty}h_n=0$ and $\lim_{n\to\infty}nh_n/\log n=\infty$.
\end{enumerate}

We obtain the following uniform rates of convergence.

\begin{theo}\label{theo2b}
In model (\ref{model}),  under the assumptions \ref{F1}, \ref{G1},
and \ref{H1}, we have
$$ \sup_{x\in[h_n, 1-h_n]}|\hat g(x)-g(x)|=O(h_n^\beta)+O_P\Big(\Big(\frac{|\log h_n|}{nh_n}\Big)^{1/\alpha}\Big).$$
\end{theo}

Note that the deterministic part $O(h_n^\beta)$ arises from
approximating the regression function by a polynomial, whereas the
random part originates from the observational error. Balancing the
two sources of error by setting $h_n \asymp ((\log
n)/n)^{\frac{1}{\alpha\beta+1}}$ gives
\begin{equation}
\label{opt-rate}
\sup_{x\in[h_n, 1-h_n]}|\hat g(x)-g(x)|=O_P\Big(\Big(\frac{\log n}{n}\Big)^{\frac{\beta}{\alpha\beta+1}}\Big).
\end{equation}
(Here $a_n\asymp b_n$ means that $0<\liminf_{n\to\infty} |a_n/b_n|\le \limsup_{n\to\infty} |a_n/b_n|<\infty$.)

This result is of particular interest in the case of irregular
error distributions, i.\,e.\ $\alpha\in (0,2)$, when the rate
improves upon the typical optimal rate $O_P(((\log
n)/n)^{\frac{\beta}{2\beta+1}})$ for estimating mean regression
functions in models with regular errors.

\begin{rem}\rm
Jirak et al.\ (2014) consider a similar boundary regression
estimator while replacing the integral in (\ref{hat-g}) by its
Riemann approximation $\sum_{i=1}^n p(\textstyle{\frac
in})I\{|\textstyle{\frac in}-x|\leq h_n\}$. In particular, they use
the Lepski method to construct a data-driven bandwidth that
satisfies $h_n \asymp ((\log n)/n)^{\frac{1}{\alpha\beta+1}}$ in
probability. For this modified estimator, we obtain the same uniform
rate of convergence as in Theorem \ref{theo2b} by replacing
Proposition \ref{prop:estbound} in the proof of Theorem \ref{theo2b}
by Theorem 3.1 in Jirak et al.\ (2014). $\blacksquare$
\end{rem}

\begin{rem}\rm
For H\"{o}lder continuous regression functions with exponent $\beta\in (0,1]$ the estimator reduces to a local maximum, i.\,e.\
$\hat g(x)=\max\{Y_i\mid i=1,\dots,n \mbox{ s.\,t.\ } |\frac in -x|\leq h_n\}$.
In this case we obtain the rate of convergence as given in Theorem \ref{theo2b}  uniformly over the whole unit interval.
$\blacksquare$
\end{rem}

\begin{figure}[h]
\hspace*{3cm}\epsfig{file=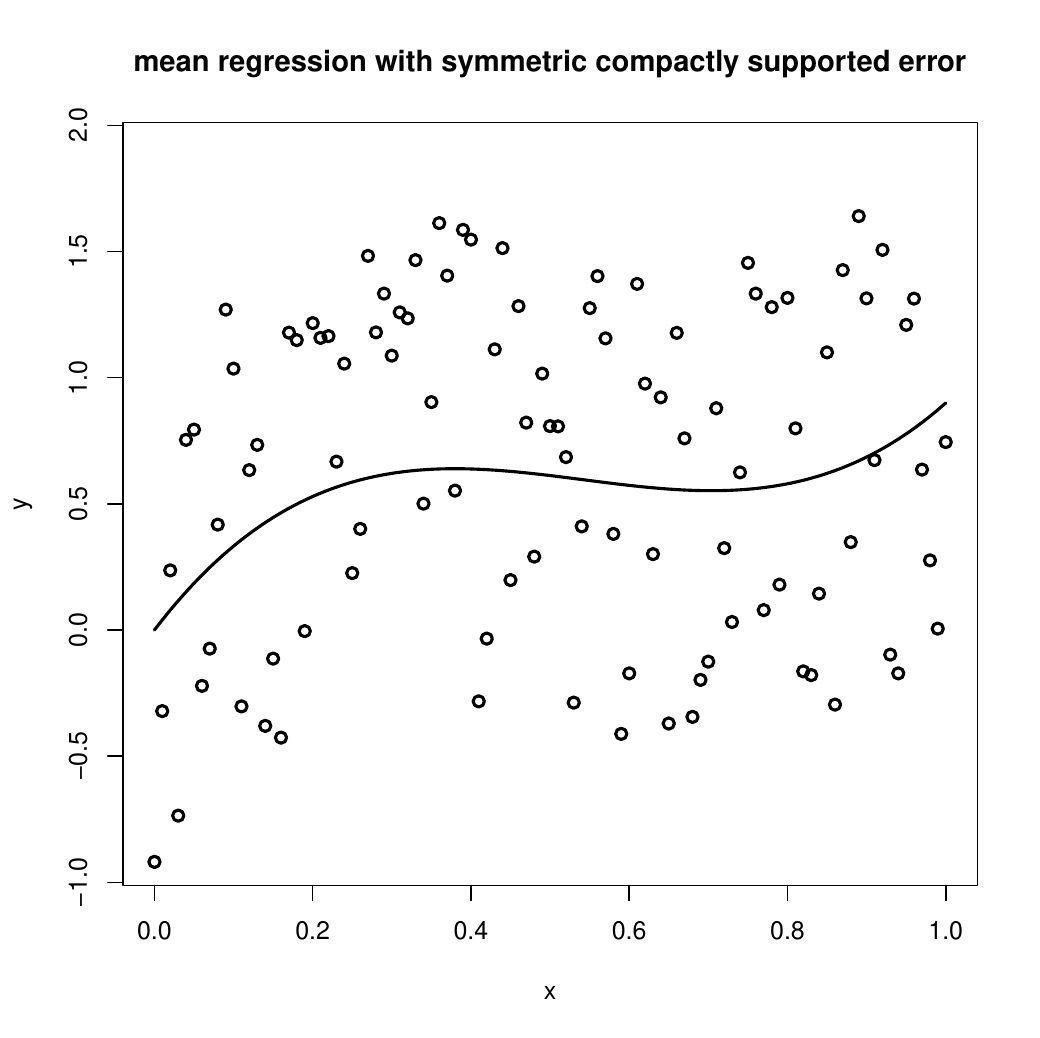,width=5cm}\hspace{1cm}\epsfig{file=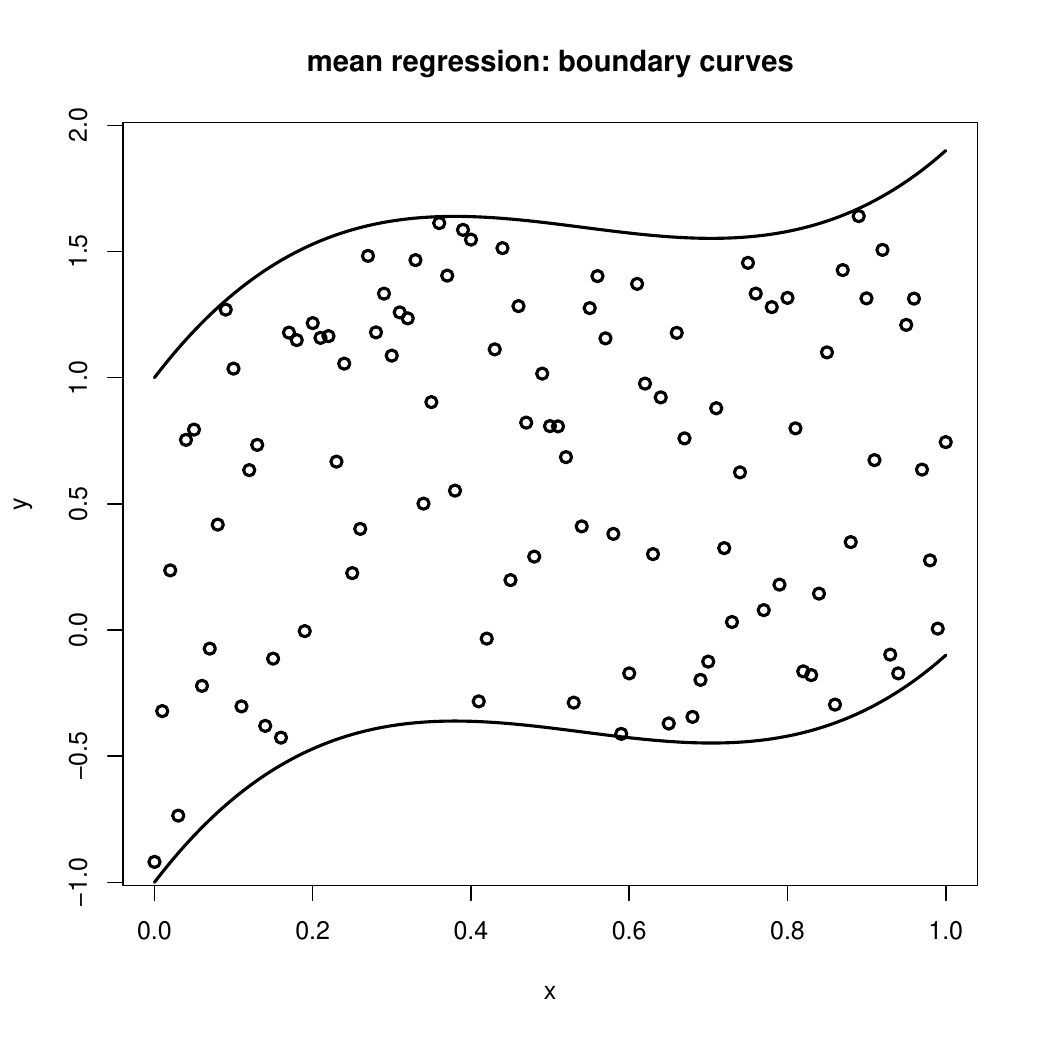,width=5cm}
\caption{\label{fig-mw}\it Example for data as in Remark \ref{mw}.}
\end{figure}

\begin{rem}\label{mw}\label{mw-lin}\rm
M\"{u}ller and Wefelmeyer (2010) consider a mean  regression model
$Y_i=m(X_i)+\eta_i$, $i=1,\dots,n,$ with symmetric error
distribution supported on $[-a,a]$ (with $a$ unknown); see the left
panel of Figure \ref{fig-mw}. The error distribution function
fulfills $F(a-y)\sim 1-y^\alpha$ for $y\searrow 0$. The local
empirical midrange of responses, i.\,e.\
$$\hat m(x)=\frac12\Big(\min_{i\in\{1,\dots,n\}\atop |X_i -x|\leq h_n}Y_i+\max_{i\in\{1,\dots,n\}\atop |X_i -x|\leq h_n}Y_i\Big)$$
is shown to have pointwise rate of convergence
$O(h_n^\beta)+O_P((nh_n)^{-1/\alpha})$ to $m(x)$ if $m$ is H\"{o}lder
continuous with exponent $\beta\in (0,1]$. Theorem \ref{theo2b}
enables us to extend M\"{u}ller's and Wefelmeyer's  (2010) results in
two ways (in a model with fixed design $X_i=\textstyle{\frac in}$):
we consider more general H\"{o}lder classes with general index $\beta
>0$, and we obtain uniform rates of convergence. To this end, we use
the mean regression estimator $\hat m=(\hat g-\hat{\tilde{g}})/2$
with $\hat g$ as before and $\hat{\tilde{g}}$ defined analogously,
but based on $(\frac in, -Y_i)$, $i=1,\dots,n$; see the right panel
of Figure \ref{fig-mw}. The rates obtained for $\sup_{x\in
[h_n,1-h_n]}|\hat m(x)-m(x)|$ are the same as in Theorem
\ref{theo2b}. $\blacksquare$
\end{rem}

\begin{rem}\rm \label{rem:HallKeilegom}
For $\beta\in (1,2]$, Hall and Van Keilegom (2009) consider the following local linear boundary regression estimator:
\begin{equation}\label{g-lin}
\check g(x)=\inf \Big\{\alpha_0\,\Big|\,
(\alpha_0,\alpha_1)\in\mathbb{R}^2: Y_i\leq
\alpha_0+\alpha_1\left(\textstyle{\frac in-x}\right)\forall i\in\{1,\dots,n\}\mbox{
s.\,t.\ } \left|\textstyle{\frac in}-x\right|\leq h_n\Big\}.
\end{equation}
Because of $\int_{x-h_n}^{x+h_h}
(\alpha_0+\alpha_1(t-x))\,dt=2\alpha_0h_n$ this estimator coincides
with $\hat g$ for $\beta\in (1,2]$. However, in the case $\beta>2$
replacing the linear  function in (\ref{g-lin}) by a polynomial of
order $\lceil\beta\rceil -1$ renders the estimator $\check g$
useless. One obtains $\check g(x)=-\infty$ for  $x\not\in\{\frac
jn\mid j=1,\dots,n\}$ while $\check g(\frac jn)=Y_j$, $j=1,\dots,n$.
This was already observed by Jirak et al.\ (2014). $\blacksquare$
\end{rem}

Note that typically the estimator $\hat g$ is not continuous.  One
might prefer to consider a smooth estimator by convoluting $\hat g$
with a kernel. Such a modified estimator will also be advantageous
when deriving an expansion for the residual based empirical
distribution function in the next section.  Therefore we define
\begin{equation}\label{tilde-g}
\tilde{g}(x)=\int_{h_n}^{1-h_n}\hat g(z)\frac 1{b_n}K\left(\frac{x-z}{b_n}\right)dz
\end{equation}
and formulate some additional assumptions.

\begin{enumerate}[label=(\textbf{K\arabic{*}})]
\item \label{K1}
$K$ is a continuous kernel with support $[-1,1]$ and order
$\lfloor\beta\rfloor +1$, i.e.\ $\int K(u)\, du=1$, $\int u^rK(u)\,
du=0$ $\forall r=1,\ldots,\lfloor\beta\rfloor$. Furthermore, $K$ is
differentiable with Lipschitz-continuous derivative $K'$ on
$(-1,1)$.
\end{enumerate}
\begin{enumerate}[label=(\textbf{B\arabic{*}})]
\item \label{B1} The sequence $(b_n)_{n\in\en}$ of positive bandwidths satisfies $\lim_{n\to\infty}b_n=0$.
\item[({\bf B2.$\boldsymbol\delta$})] 
$\displaystyle
h_n^\beta+\Big(\frac{\log n}{nh_n}\Big)^{1/\alpha}= o\big(b_n^{(1+2\delta)\vee (3-(\beta-1)(1/\delta-1))}\big)=
\left\{
\begin{array}{l@{\text{ if }}l}
o\big(b_n^{1+2\delta}\big) & \delta\leq \frac{\beta-1}{2}\\
o\big(b_n^{3-(\beta-1)(1/\delta-1)}\big) & \delta> \frac{\beta-1}{2}.
\end{array}
\right.
$
\end{enumerate}
Here we assume that the  parameter $\delta$, which quantifies the
minimal required smoothness of the estimator of $g'$, lies in
$(0,1\wedge(\beta-1))$. For example, if $\beta <3$ and the optimal
bandwidth $h_n\asymp ((\log n)/n)^{1/(\alpha\beta+1)}$ is chosen,
then ({\bf B2.$\boldsymbol\delta$}) is fulfilled with
$\delta=(\beta-1)/2$ for any $b_n$ that satisfies $h_n=o(b_n)$.

The estimator $\tilde g$ is differentiable and we obtain the following uniform rates of convergence for $\tilde g$ and its derivative $\tilde g'$.
\begin{theo}\label{theo-smooth1}
If the model assumptions (\ref{model}), \ref{F1}, \ref{G1} with
$\beta>1$, \ref{H1}, \ref{K1}, and \ref{B1} hold, then
for $I_n=[h_n+b_n,1-h_n-b_n]$
\begin{itemize}
\item[\rm(i)]
$\displaystyle \sup_{x\in I_n}|\tilde g(x)-g(x)|=O(b_n^{\beta})+O(h_n^{\beta})+O_P\Big(\Big(\frac{|\log h_n|}{nh_n}\Big)^{\frac 1\alpha}\Big)$
\item[\rm(ii)]
$\displaystyle \sup_{x\in I_n}|\tilde
g'(x)-g'(x)|=O(b_n^{\beta-1})+O\left(b_n^{-1}h_n^{\beta}\right)
+O_P\Big(b_n^{-1}\Big(\frac{|\log h_n|}{nh_n}\Big)^{\frac
1\alpha}\Big).$\\
If $h_n^\beta+(\log n/(nh_n))^{1/\alpha}=o(b_n)$, then $\sup_{x\in I_n}|\tilde
g'(x)-g'(x)|=o_P(1)$; in particular this holds if ({\bf B2.$\boldsymbol\delta$}) is
fulfilled for some $\delta\in(0,1\wedge (\beta-1))$.
\item[\rm(iii)]
For all  $\delta\in(0,1\wedge (\beta-1))$, under the
additional assumption ({\bf B2.$\boldsymbol\delta$}),
$$ \sup_{x,y\in I_n,x\neq y}\frac{|\tilde
g'(x)-g'(x)-\tilde g'(y)+g'(y)|}{|x-y|^{\delta}}=o_P(1).$$
\end{itemize}
\end{theo}

\section{The error distribution} \label{section3}

\subsection{Estimation}\label{subsection3.1}


In this section we consider estimators of the error distribution in
model (\ref{model}). For the asymptotic analysis we need the
following additional assumption.

\begin{enumerate}[label=(\textbf{F\arabic{*}})]
\setcounter{enumi}{1}
\item \label{F2} The cdf $F$ of the errors is H\"{o}lder continuous of order $\alpha\wedge 1$.
\end{enumerate}

We define residuals $\hat\eps_i=Y_i-\hat g(\frac in)$,  and a resulting modified sequential empirical distribution function by
$$\hat F_n(y,s)=\frac{1}{m_n}\sum_{i=1}^{\ns} I\{\hat\eps_i\leq y\}I\{h_n<\textstyle{\frac in}\leq 1-h_n\},$$
where $m_n=\sharp \{i\in\{1,\dots,n\}\mid h_n<\frac in\leq
1-h_n\}=n-\lfloor nh_n\rfloor -\lceil nh_n\rceil$.  We consider the
sequential process, because it will be useful for testing hypotheses
in section \ref{hyptests}. With slight abuse of notation, let $\hat
F_n(y)=\hat F_n(y,1)$ denote the corresponding estimator for $F(y)$.

We first treat a simple case where the influence of the regression
estimation on the residual  empirical process is negligible. To this
end, let $F_n$ denote the standard empirical distribution function
of the unobservable errors $\eps_1,\dots,\eps_n$. Furthermore,
define $\bar s_n=\big(\lfloor n(s\wedge(1-h_n))\rfloor-\lfloor
n(s\wedge h_n)\rfloor\big)/m_n$ and interpret $\bar s_n/\ns$ as $0$
for $s=0$.
 Note that  $\bar s_n=1$ if $s=1$ and $s_n\to s$ as $n\to\infty$, for each fixed $s$.

\begin{theo}\label{theo3a}
Assume that the conditions \ref{F1}, \ref{G1}, and \ref{F2} are
fulfilled with $\beta>1$. Furthermore, assume  $\frac
1\beta<\alpha<2-\frac{1}{\beta}$ and $h_n\asymp ((\log
n)/n)^{1/(\alpha\beta+1)}$. Then we have
$$ \sup_{y\in\mathbb{R},s\in[0,1]}|\hat F_{n}(y,s)-\bar s_nF_{\lfloor ns\rfloor}(y)|=o_P(n^{-1/2}).$$
Thus the process $\{\sqrt n(\hat{F}_{n}(y,s)-\bar s_nF(y))\mid
s\in[0,1],y\in\mathbb{R}\}$ converges  weakly to a Kiefer process
$K_F$, a centered Gaussian process with covariance function
$((s_1,y_1),(s_2,y_2))\mapsto(s_1\wedge s_2)( F(y_1\wedge
y_2)-F(y_1)F(y_2))$.
\end{theo}

\begin{rem}\rm  \label{rem:bandwidth}
 The assertion of Theorem \ref{theo3a} holds true under the following weaker conditions on the (possibly random) bandwidth:
   \begin{equation} \label{eq:alphabetacond}
       h_n=o_P\big(n^{-1/(2(\alpha\wedge 1)\beta)}\big), \quad n^{(\alpha\vee 1)/2-1}\log n = o_P(h_n).
   \end{equation}
In particular, one may use the adaptive bandwidth proposed by Jirak
et al.\ (2014).
\\
Condition \eqref{eq:alphabetacond} can be fulfilled if and only if
$\frac 1\beta<\alpha<2-\frac{1}{\beta}$, which in turn can be
satisfied for all $\alpha\in(0,2)$, provided the regression function
$g$ is sufficiently smooth. It ensures that one can choose a rate
$a_n$ of larger order than the uniform bound on the estimation error
established in Theorem \ref{theo2b} such that
      $$|F(y+a_n)-F(y)|=O(a_n^{\alpha\wedge 1})=o(n^{-1/2}). \quad \blacksquare$$
\end{rem}
\begin{rem}\rm
Theorem \ref{theo3a} implies that for $\alpha\in
(1/\beta,2-1/\beta)$ the estimation of the regression function has
no impact on the estimation of the irregular error distribution.
This is remarkably different from corresponding results on the
estimation of the error distribution in mean regression models with
regular error distributions. Here the empirical distribution
function of residuals, say $\check F_n$, is not asymptotically
$\sqrt{n}$-equivalent to the empirical distribution function of true
errors. The process  $\sqrt n(\check{F}_n-F)$ converges to a
Gaussian process whose covariance structure depends on the error distribution in a complicated way;
cf.\ Theorem 2 in Akritas and Van Keilegom (2001). In the simple
case of a mean regression model with equidistant design and an error
distribution $F$ with bounded density $f$ one has
$$\sqrt{n}(\check F_n(y)-F_n(y))=\frac{f(y)}{\sqrt{n}}\sum_{i=1}^n \eps_i+o_P(1)$$
uniformly with  respect to $y\in\mathbb{R}$ when the regression
function is estimated by a local polynomial estimator, under
appropriate bandwidth conditions (see Proposition 3 in Neumeyer and
Van Keilegom (2009)). $\blacksquare$
\end{rem}

In order to obtain asymptotic results for   estimators of the error
distribution for $\alpha\geq 2-\frac{1}{\beta}$, a finer analysis is needed.  In
what follows, we will use the smooth regression estimator
$\tilde{g}$ defined in (\ref{tilde-g}). Let $\tilde{F}_n$ denote the
empirical distribution function based on residuals
$\tilde\eps_j=Y_j-\tilde g(\frac jn)$, i.\,e.\
\[\tilde F_n(y)=\frac 1{m_n}\sjn I\{\tilde\eps_j\leq y\}I\{\textstyle{\frac jn}\in I_n\}\]
where $I_n=[h_n+b_n,1-h_n-b_n]$ and $m_n= \sharp\{j\in\{1,\dots,n\}\mid h_n+b_n\leq\frac jn\leq 1-h_n-b_n\}=n-2\lceil n(h_n+b_n)\rceil+1$.
Then the following asymptotic expansion is valid.

\begin{theo}\label{theo-smooth2}
If the conditions \ref{F1}, \ref{F2}, \ref{G1} with $\beta>1$,
\ref{H1}, \ref{K1}, \ref{B1}, and ({\bf B2.$\boldsymbol\delta$}) for
some $\delta\in (1/\alpha-1,1\wedge(\beta-1))$ are fulfilled, then
\begin{equation}  \label{eq:expan}
\tilde F_n(y) = \frac 1n\sjn I\{ \eps_j\leq y\}+\frac 1{m_n}\sjn
 \left(F\left(y+(\tilde g-g)(\textstyle{\frac jn})\right)-F(y)\right)I\{{\textstyle{\frac jn}}\in I_n\}+\op
\end{equation}
uniformly for all $y\in\er$.
\end{theo}
\begin{rem} \label{rem:withoutB2}\rm
One can choose bandwidths $h_n$ and $b_n$  such that the conditions
\ref{H1}, \ref{B1} and ({\bf B2.$\boldsymbol\delta$}) are fulfilled
for some $\delta\in (1/\alpha-1,1\wedge(\beta-1))$  if
this interval is not empty, which in turn is equivalent to $\alpha>1/(\beta\wedge 2)$.
Thus the expansion given in Theorem \ref{theo-smooth2} is also valid for regular error distributions.
\\
If one assumes  ({\bf B2.$\boldsymbol\delta$})  for some
$\delta\in(0,1\wedge(\beta-1))$, but drops the condition $\delta>1/\alpha-1$
and, in addition, replaces \ref{F2} with the assumption that $F$ is
Lipschitz continuous on $(-\infty,\kappa]$ for some $\kappa<0$, then
expansion \eqref{eq:expan} still holds uniformly on
$(-\infty,\tilde\kappa]$ for all $\tilde\kappa<\kappa$. In
particular, this holds if $F$ has a bounded density on
$(-\infty,\kappa]$. $\blacksquare$
\end{rem}

Next we examine under which conditions  the additional term in
\eqref{eq:expan} depending on the estimation error is asymptotically
negligible. We focus on those arguments $y$ which are bounded away
from 0, because in this setting weaker conditions on $\alpha$ and
$\beta$ are needed. Moreover, for the analysis of the tail behavior
of the error distribution at 0, tail empirical processes are better
suited and will be considered in future work.

Note that the estimator $\hat g$ tends to  underestimate the true
function because it is defined via a polynomial which is minimal
under the constraint that it lies above all observations
$(i/n,Y_i)$, which in turn all lie below the true boundary function.
As this systematic underestimation does not vanish from (local or
global) averaging, we first have to introduce a bias correction.

Let $E_{g\equiv 0}$ denote the expectation if the  true regression
function is identical 0. For the remaining part of this section, we
assume that $E_{g\equiv 0}(\hat g(1/2))$ is known or that it can be
estimated sufficiently accurately. For example, if the empirical
process of residuals shall be used to test a simple null hypothesis,
then one may calculate or simulate this expectation under the given
null distribution.
We define a bias corrected version of the smoothed estimator by
$$\gts(x) := \tilde g(x)-E_{g\equiv 0}(\hat g(1/2)), $$
for $x\in I_n$. The following lemma ensures that the  above results
for $\tilde g$ carry over to this variant if the following condition
on the lower tail of $F$ holds:
\begin{enumerate}[label=(\textbf{F3})]
\item \label{F3} There exists $\tau>0$ such that
$  F(-t)=o(t^{-\tau})$ as $t\to \infty$.
\end{enumerate}

\begin{lemma} \label{lem:expectg0}
  If model \eqref{model} holds with  $g$  identical 0 and the conditions \ref{F1}, \ref{F3}, \ref{G1}, and \ref{H1} are fulfilled, then for all $x\in[h_n,1-h_n]$
  $$ E_{g\equiv 0}(|\hat g_n(x)|) = E_{g\equiv 0}(|\hat g_n(1/2)|)=O\Big(\Big(\frac{\log n}{nh_n}\Big)^{1/\alpha}\Big).
  $$
\end{lemma}

We need some additional conditions on the rates at which the bandwidths $h_n$ and $b_n$ tend to 0:
\begin{enumerate}[label=(\textbf{H2})]
\item \label{H2} $h_n=o\big(n^{-1/(2\beta)}\wedge n^{-1/(\alpha\beta+1)}\big), \quad n^{\alpha/4-1}\log n=o(h_n)$
\end{enumerate}
\begin{enumerate}[label=(\textbf{B3})]
\item \label{B3} $\displaystyle b_n=o\bigg(n^{-1/(2\beta)} \wedge \Big(h_n^{-2\beta} n^{-1}\Big)\wedge\Big(\Big(\frac{nh_n}{\log n}\Big)^{2/\alpha}n^{-1}\Big) \bigg)$
\end{enumerate}

In particular, these assumptions  ensure that the bias terms of
order $h_n^\beta+b_n^\beta$ are of smaller order than $n^{-1/2}$ and
$(nh_n)^{-1/\alpha}$ and hence asymptotically negligible, and that
quadratic terms in the estimation error are uniformly negligible,
that is, $\sup_{x\in I_n}|\gts(x)-g(x)|^2=o_P(n^{-1/2})$.
\begin{theo} \label{theo:remainderterm}
 Suppose the model assumptions \eqref{model} with $\alpha\in (0,2)$, $\beta>1$, \ref{F1}, \ref{F3}, \ref{G1}, \ref{H1}, \ref{H2},
 \ref{K1}, \ref{B1}, ({\bf B2.$\boldsymbol\delta$}) for some $\delta>0$,  and \ref{B3}
 hold
   and $F$ has a bounded density on $(-\infty,\kappa]$ for some $\kappa<0$. Then
  $$ \sup_{y\in (-\infty,\kappa]}\bigg|\frac 1{m_n}\sjn \left(F\left(y+(\gts-g)(\textstyle{\frac jn})\right)-F(y)\right)
  I\{{\textstyle{\frac jn}}\in I_n\}\bigg|=o_P(n^{-1/2}).
  $$
\end{theo}
\begin{rem} \rm
 The conditions on $h_n$ and $b_n$ used in Theorem
\ref{theo:remainderterm} can be fulfilled if and only if
$\alpha<2\beta-1$. In particular, this theorem is applicable if
$\beta\geq 3/2$ and the error distribution is irregular, i.e.,
$\alpha<2$. A possible choice of bandwidths is
$$ h_n\asymp \big(n^{-1/(2\beta)}\wedge n^{-1/(\alpha\beta+1)}\big)/\log n, \quad
  b_n \asymp n^{-\lambda} \text{ for some } \lambda\in \Big(\frac 1{2\beta}, \frac\beta{\alpha\beta+1}\wedge \frac{2\beta-1}{2\alpha\beta}\Big).\quad  \blacksquare
$$
\end{rem}

We obtain asymptotic equivalence of the empirical process of
residuals (restricted to $(-\infty,\kappa]$) to the empirical
process of the errors. To formulate the result, let $\tilde F_n^*$
be defined analogously to $\tilde F_n$, but with $\tilde g$ replaced
by $\tilde g^*$.

\begin{cor}
Under the assumptions of Theorems \ref{theo-smooth2} and \ref{theo:remainderterm}, we have $
\sup_{y\in(-\infty,\kappa]}|\tilde F_n^*(y)-F_n(y)|=o_P(n^{-1/2}).$
Thus the process $(\sqrt n(\tilde F_n^*(y)-F(y)))_{y\in
(-\infty,\kappa]}$ converges weakly to a centered Gaussian process
with covariance function $(y_1,y_2)\mapsto F(y_1\wedge
y_2)-F(y_1)F(y_2)$, $y_1,y_2\in (-\infty,\kappa]$.
\end{cor}

Note that for the Corollary one needs the condition $1/(\beta\wedge 2)<\alpha<(2\beta-1)\wedge 2$.

\section{Hypotheses testing}\label{hyptests}

\subsection{Goodness-of-fit testing}\label{subsection3.2}

Let $\cf=\{F_\vartheta\mid\vartheta\in\Theta\}$ denote a continuously parametrized class of error distributions such that for each $\vartheta\in\Theta$, $F_\vartheta(y)=1-c_\vartheta|y|^{\alpha_\vartheta}+r_\vartheta(y)$
with $r_\vartheta(y)=o(|y|^{\alpha_\vartheta})$ for $y\nearrow 0$. Our aim is to test the null hypothesis
$H_0: F\in\cf.$
We assume that $\alpha_\vartheta \in (1/\beta,2-1/\beta)$ for all $\vartheta\in\Theta$, such that Theorem \ref{theo3a} can be applied  under $H_0$.
Let $\hat\vartheta$ denote an estimator for $\vartheta$ based on residuals $\hat\eps_i=Y_i-\hat g(\frac in)$, $i=1,\dots,n$. The goodness-of-fit test is based on the empirical process
$$ S_n(y)=\sqrt{n}(\hat F_n(y)-F_{\hat \vartheta}(y)), \quad y\in\mathbb{R},
$$
where, as before, $\hat F_n(y)=\hat F_n(y,1)$.
Under any fixed alternative that fulfills \ref{F1} for some $\alpha$, $\hat g$ still uniformly consistently estimates $g$, and thus $\hat F_n$ is a consistent estimator of the error distribution $F$. If $\hat\vartheta$  converges to some $\vartheta^*\in\Theta$ under the alternative, too, then a consistent hypothesis test is obtained by rejecting $H_0$ for large values of, e.\,g., a Kolmogorov-Smirnov test statistic $\sup_{y\in\mathbb{R}}|S_n(y)|$.
Note that under $H_0$ it follows  from Theorem \ref{theo3a} that
$$S_n(y)=\sqrt{n}(F_n(y)-F_\vartheta(y))-\sqrt{n}(F_{\hat\vartheta}(y)-F_\vartheta(y))+o_P(1),$$
where $\vartheta$ denotes the true parameter. We consider two examples.

\begin{example}\rm\label{ex-uniform} Consider the  mean regression model $Y_i=m(\frac in)+\eta_i$, $i=1,\dots,n,$ with symmetric error cdf $F$ and $\beta >1$, and define $\hat m$ with some bandwidth $h_n\asymp ((\log n)/n)^{1/(\alpha\beta+1)}$  as in Remark \ref{mw-lin}.
We want to test the null hypothesis $H_0: F\in\cf=\{F_\vartheta\mid\vartheta\in\Theta\}$ for some $\Theta\subset (0,\infty)$, where $F_\vartheta$ denotes the distribution function of the uniform distribution on $[-\vartheta,\vartheta]$ (with $\alpha_\vartheta=1$ for all $\vartheta>0$). Define residuals $\hat\eta_i=Y_i-\hat m (\frac in)$, $i=1,\dots,n$, and let
$$\hat\vartheta_n=\max\Big(\max_{nh_n\leq i\leq n-nh_n}\hat\eta_i,{-\!\!\!\!\min_{nh_n\leq i\leq n-nh_n}\hat\eta_i}\Big)
=\max_{nh_n\leq i\leq n-nh_n}|\hat\eta_i|.$$
Then $|\hat\vartheta_n-\vartheta|$ is bounded by $|\max_{nh_n\leq i\leq n-nh_n} |\eta_i|-\vartheta|+\sup_{x\in[h_n,1-h_n]}|\hat m(x)-m(x)|
=o_P(n^{-1/2})$.
Since $F_\vartheta(y)=\frac{y+\vartheta}{2\vartheta}I_{[-\vartheta,\vartheta]}(y)
+I_{(\vartheta,\infty)}(y)$, one may conclude $\sup_{y\in\er}|F_{\hat\vartheta_n}(y)-F_\vartheta(y)|=o_P(n^{-1/2})$. Thus the process $S_n$ converges weakly to a Brownian bridge $B$ composed with $F$. The Kolmogorov-Smirnov test statistic $\sup_{y\in\mathbb{R}}|S_n(y)|$ converges in distribution to $\sup_{t\in [0,1]}|B(t)|$. Thus although our testing problem requires the estimation of a nonparametric function and we have a composite null hypothesis,  the same asymptotic distribution arises as in the Kolmogorov-Smirnov test for the simple hypothesis $H_0: F=F_0$ based on an iid sample with distribution $F$.
$\blacksquare$
\end{example}

\begin{example}\rm \label{ex-weibull} Again assume that the H\"{o}lder coefficient $\beta$ is greater than 1. Consider the null hypothesis $H_0: F\in\cf=\{F_{\vartheta}\mid \vartheta\in (0,\infty)\}$, where $F_{\vartheta}(y)=e^{-(-\vartheta y)^{\alpha}}I_{(-\infty,0)}(y)+I_{[0,\infty)}(y)$ denotes a Weibull distribution with some fixed shape parameter $\alpha\in (1/\beta,2-1/\beta)$ and unknown scale parameter $\thet$.
Note that $F_\thet$ satisfies \ref{F1} with $c=\thet$.

Define the moment estimator $\hat\vartheta_n=\left(\frac 1{m_n}\sjn(-\he_j)^\alpha I\{h_n<\frac jn\leq 1-h_n\}\right)^{-\frac 1{\alpha}}$ which is motivated by $E_{\vartheta}[(-\e_1)^\alpha]=\vartheta^{-\alpha}$. A Taylor expansion of $x\mapsto x^{\alpha}$ at $x=-\eps_j$ yields
\beq
\hat\vartheta_n^\alpha-\vartheta^\alpha&=&
-(\hat\vartheta_n\vartheta)^\alpha\Big(\frac 1{m_n}\sjn((-\e_j)^\alpha-\thet^{-\alpha})I\{h_n<\frac jn\leq 1-h_n\}\\
&&\qquad\qquad+\frac \alpha{m_n}\sjn(-\xi_j)^{\alpha-1}\Big(\hat g(\frac jn)- g(\frac jn)\Big)I\{h_n<\frac jn\leq 1-h_n\}\Big)\\
 & = & -\vartheta^{2\alpha}\frac 1{m_n}\sjn((-\e_j)^\alpha-\vartheta^{-\alpha})I\{h_n<\frac jn\leq 1-h_n\}+o_{P_\vartheta}(n^{-1/2})\\
 & = & O_{P_\vartheta}(n^{-1/2})
\eeq
for some $\xi_{j}$ between $\he_j$ and $\e_j$, where in the last steps we have applied Theorem \ref{theo2b}, the law of large numbers and a central limit theorem.

For all $z,\tilde z\in\er$ one has $|e^{-z}-e^{-\tilde z}-(z-\tilde z)e^{-z}|=e^{-z}|e^{z-\tilde z}-1-(z-\tilde z)|\le e^{-z\wedge\tilde z}(z-\tilde z)^2.$
Thus
$$
\big|F_{\hat\vartheta_n}(y)-F_\thet(y)-e^{-(-\vartheta y )^\alpha}\big(\hat\vartheta_n^\alpha-\vartheta^\alpha\big)^2|  \le e^{-(-(\hat\vartheta_n\wedge\thet)y)^\alpha} \big((\hat\thet_n^\alpha-\thet^\alpha)(-y)^\alpha\big)^2
 =  O_{P_\vartheta}(n^{-1})
$$
uniformly for all $y\in (-\infty,0]$. Now analogously to the proof of Theorem 19.23 in van der Vaart (2000) we can conclude weak convergence of
\begin{eqnarray*}
S_n(y)&=&\sqrt n(F_n(y)-F(y))-e^{-(-\vartheta y )^\alpha}\vartheta^{2\alpha}(-y)^\alpha\frac {\sqrt n}{m_n}\sjn((-\e_j)^\alpha-\frac 1{\vartheta^\alpha})I\{h_n<\frac jn\leq 1-h_n\}\\
&&{}+o_{P_{\vartheta}}(1),
\end{eqnarray*}
$y\in\er$, to a Gaussian process with covariance function
$(y_1,y_2)\mapsto F_{\vartheta}(y_1\wedge y_2)-F_{\vartheta}(y_1)F_{\vartheta}(y_2)-e^{-(-\vartheta)^{\alpha}(y_1^\alpha+y_2^\alpha)}(y_1y_2)^\alpha \vartheta^{2\alpha}$, where the covariance function follows by simple calculations and the fact that $E_{\vartheta}[I\{\e_1\leq y\}((-\e_1)^\alpha-\vartheta^{-\alpha})]=(-y)^\alpha e^{-(-\vartheta y)^\alpha}$.

For the special case of a test for exponentially distributed errors ($\alpha=1$), the asymptotic quantiles for the Cram\'{e}r-von-Mises test statistic $\int S_n(y)^2dF_{\hat\vartheta_n}(y)$ are tabled in Stephens (1976).
$\blacksquare$
\end{example}

\subsubsection*{ Simulations}

To study the finite sample performance of our goodness-of-fit test, we investigate its behaviour on simulated data according to Examples \ref{ex-uniform} and \ref{ex-weibull} for samples of size $50, 100, 200$ and $500$. In both settings the regression function is given by $g(x)=0.5\sin(2\pi x)+4x$. We use the local linear estimator (corresponding to $\beta=2$) with bandwidth $n^{-\frac 13}$, which is up to a log term of optimal rate for $\alpha=1$ and $\beta=2$. The hypothesis tests are based on the adjusted Cram\'{e}r-von-Mises test statistic $\frac {m_n} n\int S_n(y)^2dF_{\hat\vartheta_n}(y)$ and have nominal size $5\%$. The results reported below are based on
$200$ Monte Carlo simulations for each model.

\begin{figure}[h!]\begin{center}
\begin{minipage}[t]{0.9\textwidth}
\includegraphics[width=\textwidth]{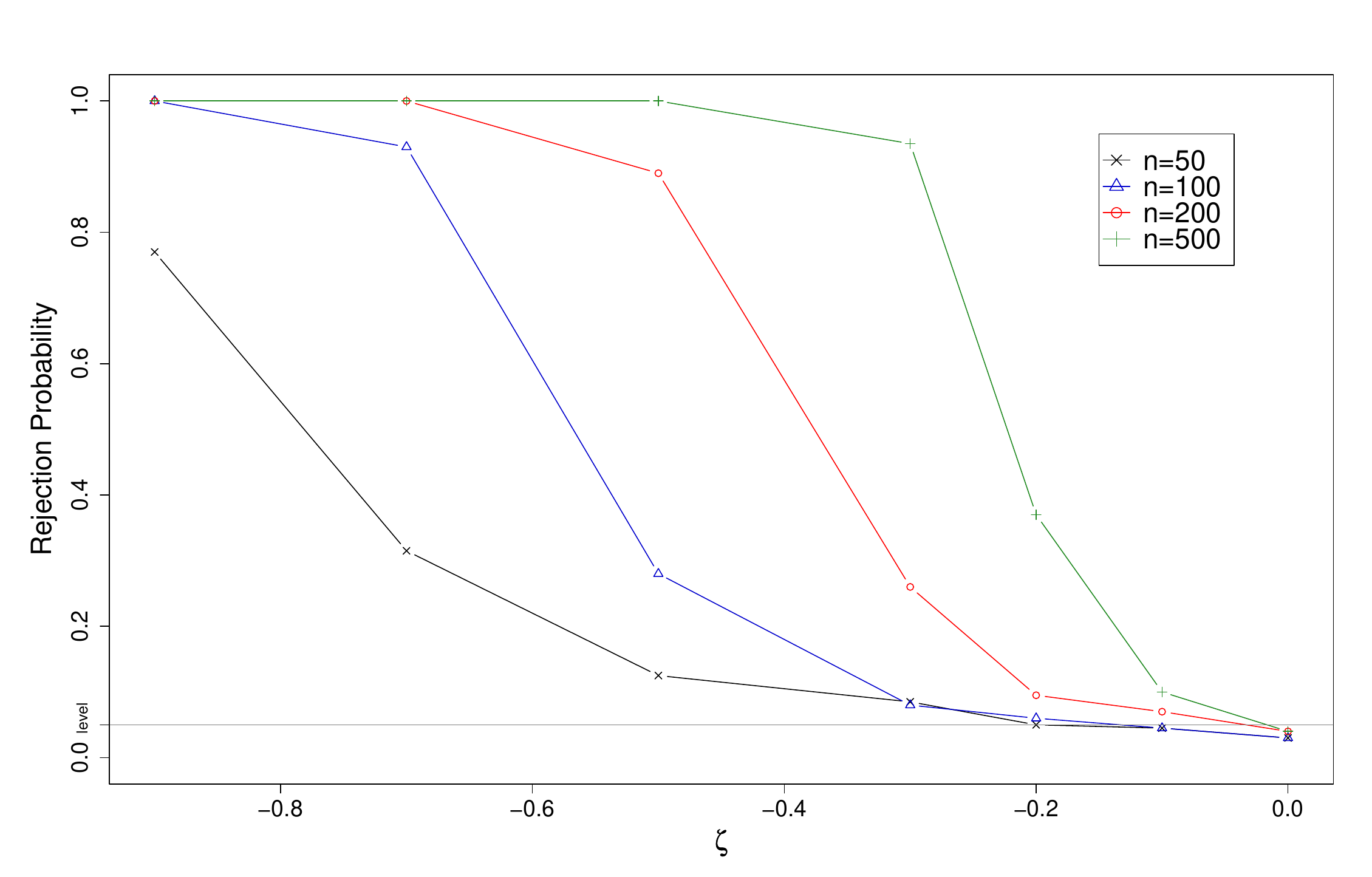}
\end{minipage}
\caption{Monte-Carlo simulations for Example \ref{ex-uniform}}\label{fig-uniform}
\end{center}
\end{figure}

In the situation of Example \ref{ex-uniform}, the errors are drawn according to the density  $f_\eps(y)=0.5(\zeta+1)(1-|y|)^\zeta I_{[-1,1]}(y)$  for different values of $\zeta\in (-1,0]$ . Note that the null hypothesis $H_0:\exists \vartheta:\e_i\sim U[-\vartheta,\vartheta]$ holds if and only if $\zeta=0$. Figure \ref{fig-uniform} shows the empirical power of the Cram\'{e}r-von-Mises type test. The actual size is close to the nominal level for all sample sizes and the power function is monotone both in $\zeta$ and the sample size $n$. For parameter values $\zeta\in[ -0.2$,0), one needs rather large sample sizes to detect the alternative, as the error distribution is too similar to the uniform distribution.

\begin{figure}[h!]\begin{center}
\begin{minipage}[t]{0.9\textwidth}
\includegraphics[width=\textwidth]{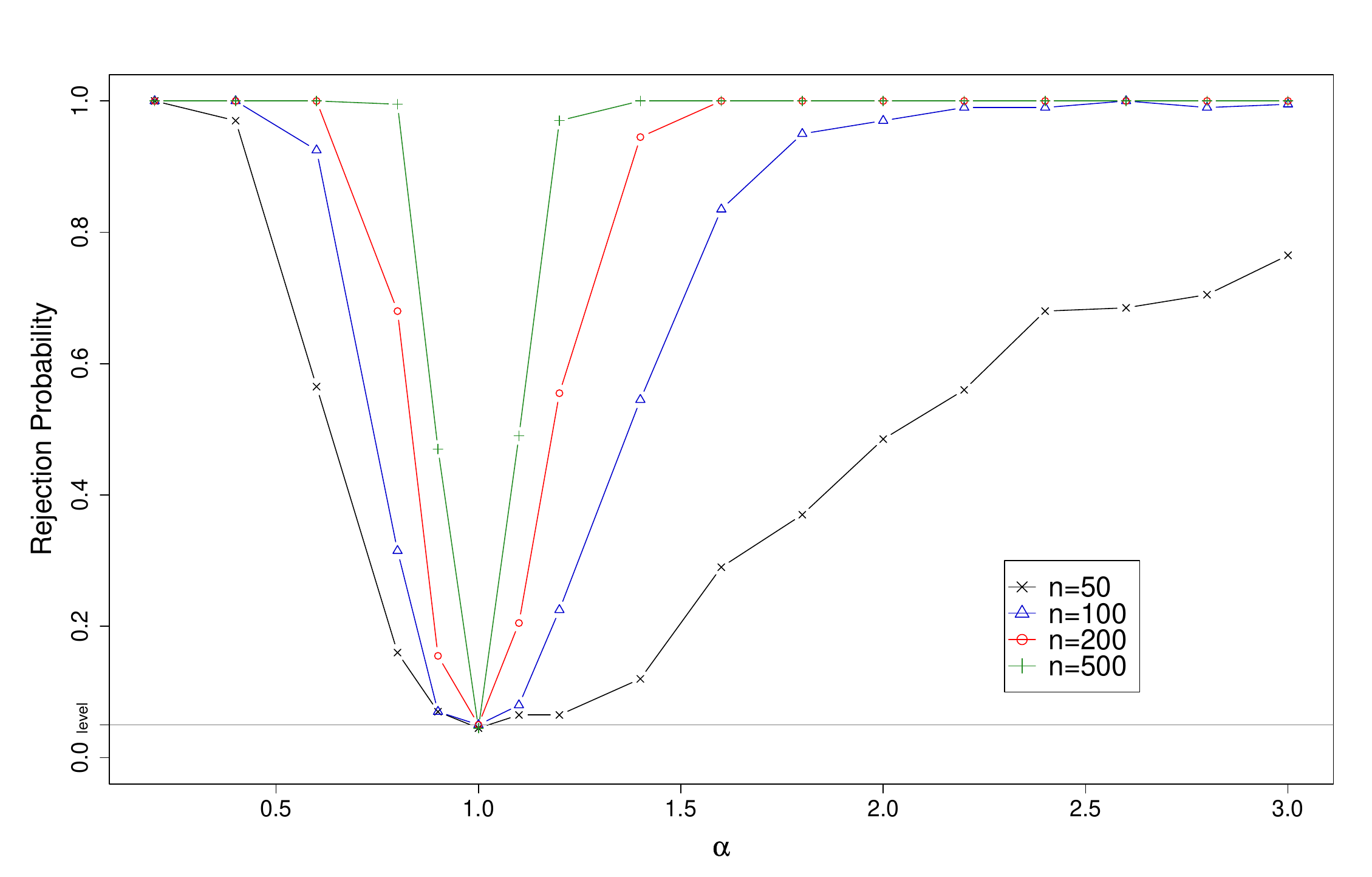}
\end{minipage}
\caption{Monte-Carlo simulations for Example \ref{ex-weibull}}\label{fig-exponential}
\end{center}
\end{figure}

In the setting of Example \ref{ex-weibull} we simulate Weibull$(\vartheta,\alpha)$ distributed errors for $\vartheta=1$ and different values of $\alpha>0$. We test the null hypothesis $H_0:\exists\vartheta:-\e_i\sim Exp(\vartheta)$ of exponentiality,  which is  only fulfilled for $\alpha=1$. In Figure \ref{fig-exponential} the empirical power function of our test is displayed for different sample sizes. Again the actual size is close to the $5\%$ and the power increases with $\alpha$ departing from one as well as with increasing $n$.

To examine the influence of the bandwidth choice, in addition we have simulated the same models with $h_n=c\cdot n^{-\frac 13}$ for different values of $c$ ranging from $c=0.2$ to $c=1.2$. The results for the test of uniformity in Example \ref{ex-uniform}  are similar to those displayed in Figure \ref{fig-uniform} for all these bandwidths. In the situation of Example \ref{ex-weibull} we obtain similar   power functions as reported above for $c$ between $0.8$ and $1.2$, whereas for smaller bandwidths the actual size of the test exceeds its nominal value substantially.

\subsection{Test for independence} \label{subsect:test_indep}

In model \eqref{model} we assume that the distributions of the errors $\eps_i$ ($i=1,\dots,n$) do not depend on the point of measurement $x_i=i/n$. We can test this assumption by comparing the sequential empirical distribution function $\hat F_n(y,s)$ for the residuals with the estimator $\bar s_n\hat F_n(y)$, which should behave similarly if the errors are iid.
The following corollary to Theorem \ref{theo3a} describes the asymptotic behavior of the
 Kolmogorov-Smirnov type test statistic
$$T_n=\sup_{s\in[0,1], y\in\mathbb{R}}\sqrt{n}|\hat F_{n}(y,s)-\bar s_n\hat F_n(y)|$$
under the null hypothesis of iid errors.

\begin{cor}\label{cor-indep}
Assume model (\ref{model}) with  \ref{F1}, \ref{F2}, \ref{G1}, and  $\frac 1\beta<\alpha<2-\frac{1}{\beta}$. Choose a bandwidth $h_n\asymp((\log n)/n)^{1/(\alpha\beta+1)}$.

Then $T_n$ converges in distribution to $\sup_{s\in[0,1],z\in[0,1]}|G(s,z)|$ where $G$ is a completely tucked Brownian sheet, i.\,e.\ a centered Gaussian process with covariance function \linebreak $((s_1,z_1),(s_2,z_2))\mapsto (s_1\wedge s_2-s_1s_2)(z_1\wedge z_2-z_1z_2)$.
\end{cor}
The proof is given in the appendix. Note that under the assumptions of the corollary the limit of the test statistic $T_n$ is distribution free. The asymptotic quantiles tabled by Picard (1985) can be used to determine the critical value for a given asymptotic size of the test.

\subsection{Test for monotone boundary functions}

\begin{figure}[h]

\centerline{\epsfig{file=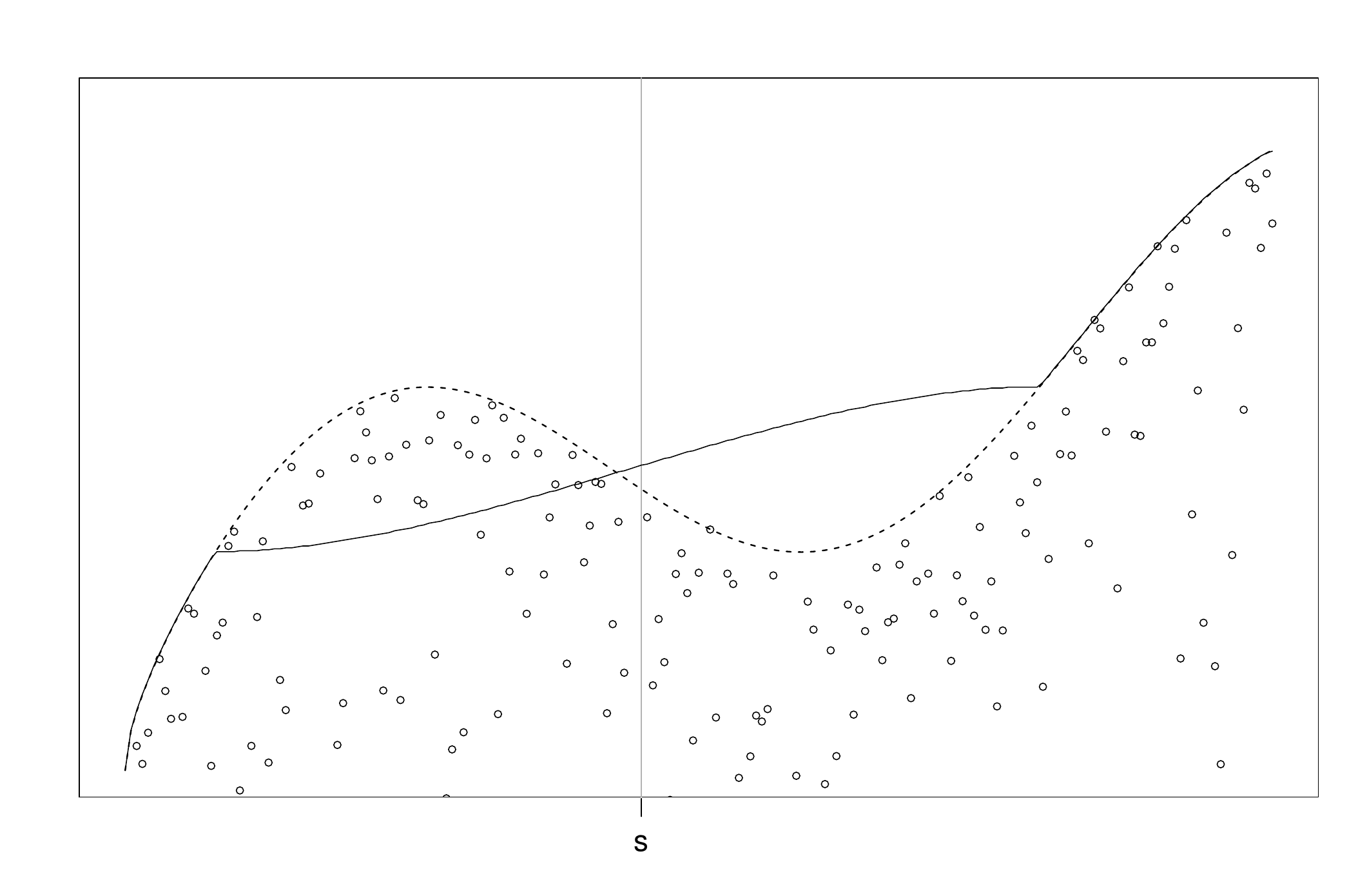,width=10cm}}

\centerline{\epsfig{file=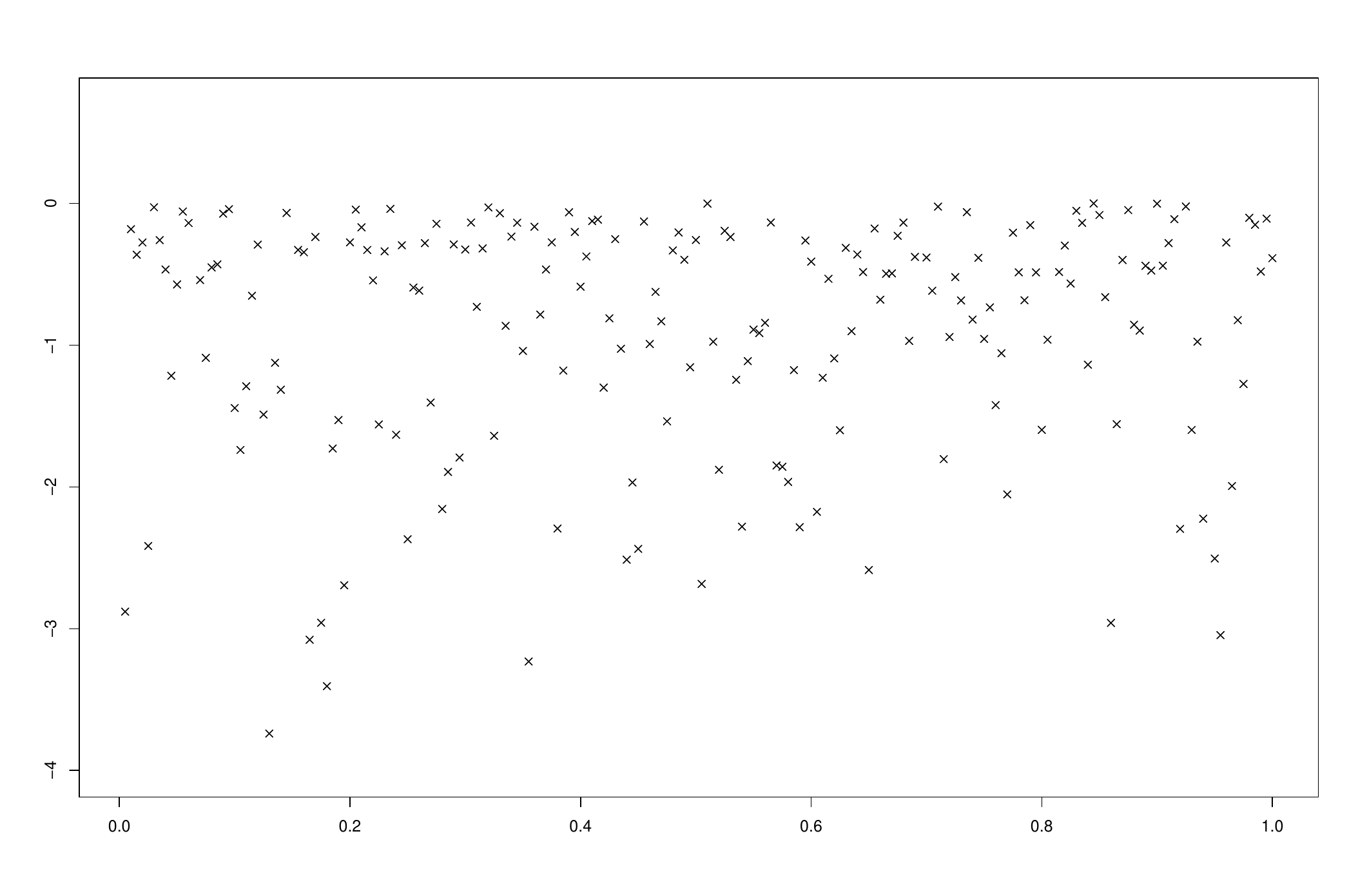,width=5cm}\epsfig{file=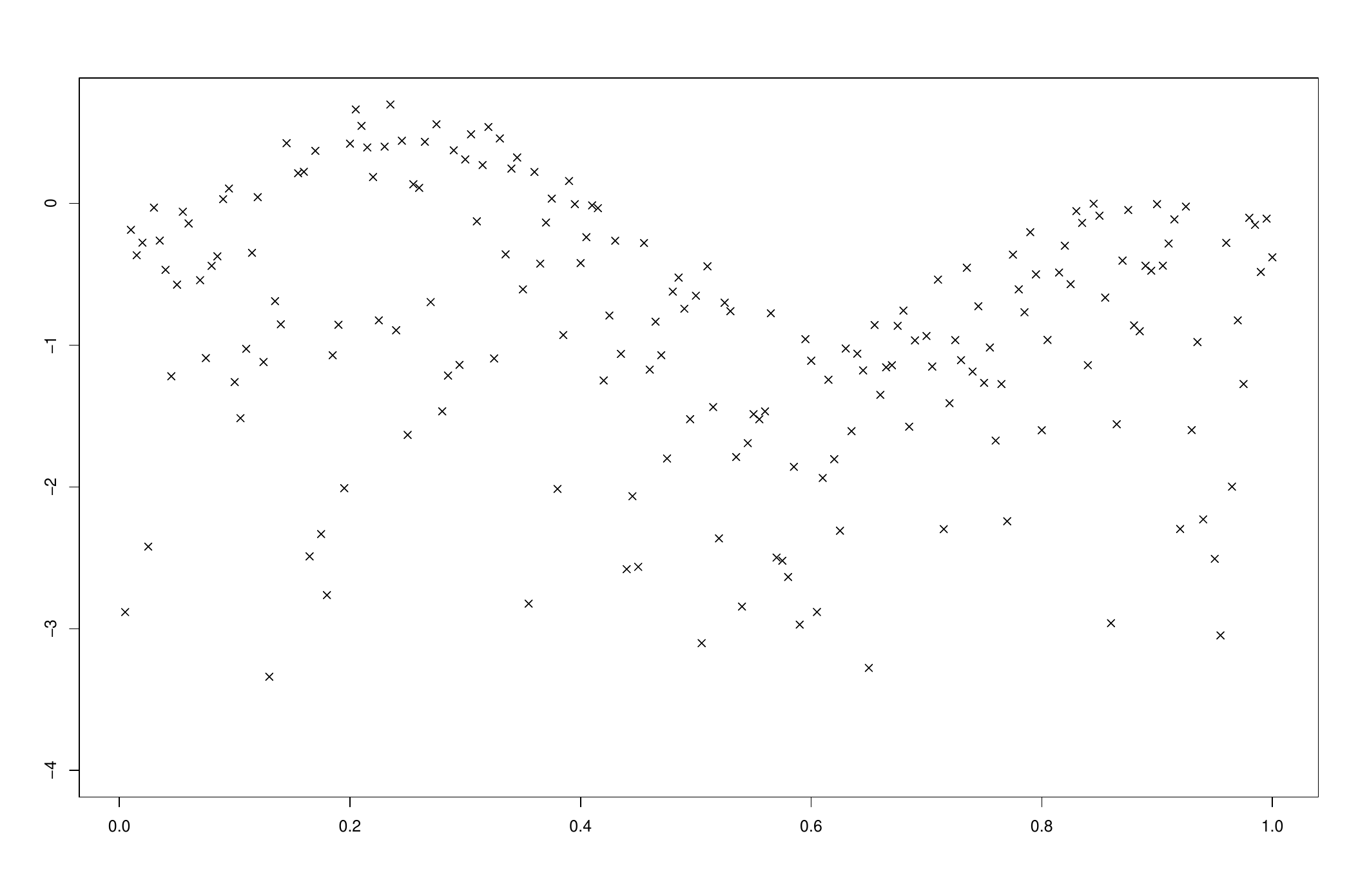,width=5cm}}
\caption{\label{monotone-graphic}\it The upper panel shows data points and the true boundary function (dashed curve) as well as the increasing rearrangement (solid curve).
The lower left panel shows the errors. The lower right panel shows residuals built with respect to the increasing rearrangement.}
\end{figure}

We consider model (\ref{model}) and aim at testing the null hypothesis
$$H_0: \quad g \mbox{ is increasing},$$
which is a common assumption in boundary models.
Let $\tilde g$ denote the smooth local polynomial estimator for $g$ defined in \eqref{tilde-g}. Such an unconstrained estimator can be modified to obtain an increasing estimator $\tilde g_I$.
To this end, for any function $h:[0,1]\to\R$ define the increasing rearrangement on $[a,b]\subset [0,1]$ as the function $\Gamma(h):[a,b]\to\R$ with
\begin{eqnarray*}
\Gamma(h)(x) &=& \inf\Big\{z\in\R\;\Big| a + \int_{a}^{b} I\{h(t)\leq z\}\, dt\geq x\Big\}.
\end{eqnarray*}
Denote by $\Gamma_n$ the operator $\Gamma$ with $[a,b]=I_n$.
We define the increasing rearrangement of $\tilde g$ as $\tilde g_I=\Gamma_n(\tilde g)$, so that $\tilde g_I=\tilde g$ if $\tilde g$ is nondecreasing (see Anevski and Foug\`{e}res, 2007, and Chernozhukov et al., 2009).  We now consider  residuals obtained from the monotone estimator: $\hat\eps_{I,i}=Y_i-\tilde g_I(\textstyle{\frac in})$, $i=1,\dots,n$. Under the null hypothesis, these residuals should be approximately iid, whereas under the alternative they show a varying behavior for $\frac in$ in different subintervals of $[0,1]$. For illustration see Figure \ref{monotone-graphic} where we have generated a data set (upper panel) with true non-monotone boundary curve $g$ (dashed curve). The solid curve is the increasing rearrangement $g_I$. The lower left panel shows the errors $\eps_i$, $i=1,\dots,n$, with iid-behaviour. The lower right panel shows $\eps_{I,i}=Y_i-g_I(\textstyle{\frac in})$, $i=1,\dots,n$, with a clear non-iid pattern.

Similarly as in Subsection \ref{subsect:test_indep}, we compare the sequential empirical distribution function
\[\tilde F_{I,n}(y,s)=\frac 1{m_n}\sjns I\{\tilde\eps_{I,j}\leq y\}I\{\textstyle{\frac jn}\in I_n\}\]
based on the increasing estimator $\tilde g_I$ with the product estimator $\bar s_n \tilde F_n(y,1)$, where again $I_n:=[h_n+b_n,1-h_n-b_n]$ and $m_n:= n-2\lceil n(h_n+b_n)\rceil+1$. Let
$$ \tilde G_n(s,y)=\sqrt{n}(\tilde F_{I,n}(y,s)- \bar s_n\tilde F_{n}(y)), \quad s\in [0,1], y\in\mathbb{R}. $$
To derive its limit distribution under the null hypothesis, we need an additional assumption:
\begin{enumerate}[label=(\textbf{I1})]
\item \label{I1} Let $\inf_{x\in[0,1]} g'(x)>0$.
\end{enumerate}

\begin{theo}\label{theo-mon}
Assume model (\ref{model}) with \ref{F1}, \ref{F2}, \ref{G1}, \ref{K1}, \ref{I1}, $\beta>1$ and $\frac 1\beta<\alpha<2-\frac{1}{\beta}$. If $h_n\asymp((\log n)/n)^{1/(\alpha\beta+1)}$ and $b_n\asymp((\log n)/n)^{1/(\alpha\beta+1)}$, then
\begin{equation}  \label{eq:tildeFconv}
 \sup_{y\in\mathbb{R},s\in[0,1]}|\tilde F_{I, n}(y,s)-\bar s_nF_{\lfloor ns\rfloor}(y)|=o_P(n^{-1/2}).
\end{equation}
Thus the Kolmogorov-Smirnov test statistic $\sup_{s\in[0,1],y\in\er}|\tilde G_n(s,y)|$, converges in distribution to $\sup_{s\in[0,1],z\in[0,1]}|G(s,z)|$ where $G$ is the completely tucked Brownian sheet (see Corollary \ref{cor-indep}).
\end{theo}
 The conditions on the bandwidths can be substantially relaxed; cf.\ Remark \ref{rem:bandwidth}.

\begin{rem}\rm
A test that rejects $H_0$ for large values of the Kolmogorov-Smirnov test statistic $T_n=\sup_{s\in[0,1],y\in\mathbb{R}}|\tilde G_n(s,y)|$ is consistent. To see this note that by Theorem 1 of Anevski and Foug\`{e}res (2007), $\sup_{x\in I_n}|\tilde g_I(x)-g_I(X)|\le \sup_{x\in I_n}|\tilde g(x)-g(x)|=o_P(1)$ with $g_I$ denoting the increasing rearrangement of $g$. Thus $n^{-1/2}T_n$ converges to
$$T=\sup_{s\in [0,1],y\in\mathbb{R}}\Big|\int_0^s F_\eps(y+(g_I-g)(x))\,dx-sF_\eps(y)\Big|.$$
Since $T>0$ under the alternative hypothesis  $g\ne g_I$,  the test statistic $T_n$ converges to infinity.
$\blacksquare$
\end{rem}

\begin{appendix}

\section{Appendix: Proofs}

\subsection{Auxiliary results}

\begin{prop} \label{prop:estbound}
  Assume that model (2.1) holds and that the regression function $g$ fulfills condition \ref{G1} for some $\beta\in(0,\beta^*]$ and some $c_g\in[0,c^*]$. Then there exist constants\; $L_{\beta^*,c^*},L_{\beta^*}>0$  and a natural number $j_{\beta^*}$ (depending only on the respective subscripts) such that
  $$|\hat g(x)-g(x)|\le  L_{\beta^*,c^*}h_n^\beta+L_{\beta^*}\max_{1\le j\le 2j_{\beta^*}} \big(\min_{i:-1+(j-1)/j_{\beta^*}\le|i/n-x|/h_n\le -1+j/j_{\beta^*} }|\eps_i|\big).
  $$
\end{prop}

This proposition can be verified by an obvious modification of the proof of Theorem 3.1 by Jirak et al.\ (2014).

\begin{lemma}\label{lem1}
Under assumptions \ref{F1} and \ref{H1} for any fixed set $I_1,\ldots, I_m$ of disjoint non-degenerate subintervals of $[-1,1]$ we have
$$\sup_{x\in [h_n,1-h_n]}\max_{1\le j\le m}
\min_{i\in\{1,\dots,n\}\atop (i/n-x)/h_n\in I_j} |\eps_i|=O_P\Big(\Big(\frac{|\log h_n|}{nh_n}\Big)^{1/\alpha}\Big).$$
\end{lemma}

\noindent
{\bf Proof.}
Let $r_n := \big(|\log h_n|/(nh_n)\big)^{1/\alpha}$.
Obviously it suffices to prove that for all non-degenerate subintervals $I\subset[-1,1]$ there exists a constant $L$ such that
$$ \lim_{n\to\infty} P\Big\{ \sup_{x\in[h_n,1-h_n]} \min_{i\in\{1,\ldots,n\} \atop (i/n-x)/h_n\in I} |\eps_i|>L r_n\Big\} = 0.
$$
Denote by $d=\sup I-\inf I>0$ the diameter of $I$ and let $d_n:=\ceil{nh_n d}-1$ and $l_n:=\floor{n/d_n}$. Then for all $x>0$
\begin{eqnarray*}
  P\Big\{ \sup_{x\in[h_n,1-h_n]} \min_{i\in\{1,\ldots,n\} \atop (i/n-x)/h_n\in I} |\eps_i|>x\Big\}
  & \le & P\Big\{ \max_{j\in\{1,\ldots,n-d_n\}} \min_{i\in\{j,\ldots,j+d_n\}} |\eps_i|>x\Big\} \\
  & \le & P\Big\{ \max_{l\in\{0,\ldots,l_n\} \atop l \text{ even}} M_{n,l}>x\Big\} + P\Big\{ \max_{l\in\{0,\ldots,l_n\} \atop l \text{ odd}} M_{n,l}>x\Big\}
\end{eqnarray*}
with
$$ M_{n,l} := \max_{j\in\{ld_n+1,\ldots,(l+1)d_n\}} \min_{i\in\{j,\ldots,j+d_n\}} |\eps_i|.
$$
Since the random variables $M_{n,l}$ for $l$ even are iid, we have
$$ P\Big\{ \max_{l\in\{0,\ldots,l_n\} \atop l \text{ even}} M_{n,l}>x\Big\} = 1-\big(1-P\{M_{n,0}>x\}\big)^{\floor{l_n/2}+1},
$$
and an analogous equation holds for the maxima over the odd numbered block maxima $M_{n,l}$.

Let $G$ be the cdf of $|\eps_i|$. If $M_{n,0}$ exceeds $x$, then there is a smallest index $j\in\{1,\ldots,d_n\}$ for which $\min_{i\in\{j,\ldots,j+d_n\}} |\eps_i|>x$. Hence
\begin{eqnarray*}
  P\{M_{n,0}>x\} & = & P\Big\{\min_{i\in\{1,\ldots,1+d_n\}} |\eps_i|>x\Big\} + \sum_{j=2}^{d_n} P\Big\{|\eps_{j-1}|\le x, \min_{i\in\{j,\ldots,j+d_n\}} |\eps_i|>x\Big\}\\
  & = & (1-G(x))^{d_n+1} + (d_n-1)G(x)(1-G(x))^{d_n+1}\\
  & \le & (1+d_nG(x))(1-G(x))^{d_n}.
\end{eqnarray*}
To sum up, we have shown that
$$ P\Big\{ \sup_{x\in[h_n,1-h_n]} \min_{i\in\{1,\ldots,n\} \atop (i/n-x)/h_n\in I} |\eps_i|>Lr_n\Big\}  \le 2\bigg(1-\Big(1-(1+d_nG(Lr_n))(1-G(Lr_n))^{d_n}\Big)^{\floor{l_n/2}+1}\bigg).
$$
It remains to be shown that the right hand side tends to 0 for sufficiently large $L$ which is true if and only if
$$ (1+d_nG(Lr_n))(1-G(Lr_n))^{d_n} = o(1/l_n).$$
This is an immediate consequence of $1/l_n\sim dh_n$ and
\begin{eqnarray*}
 \lefteqn{G(Lr_n)= cL^\alpha \frac{|\log h_n|}{nh_n} (1+o(1))}\\
 & \Longrightarrow  & (1-G(Lr_n))^{d_n}=\exp\Big(-nh_ndcL^\alpha \frac{|\log h_n|}{nh_n}(1+o(1))\Big) \\
 & \Longrightarrow  & (1+d_nG(Lr_n))(1-G(Lr_n))^{d_n} = O\Big(|\log h_n|\exp\big(-cdL^\alpha|\log h_n|(1+o(1))\big)\Big)=o(h_n)
\end{eqnarray*}
if $cdL^\alpha>1$.
\hfill $\Box$

\subsection{Proof of Theorem \ref{theo2b}}
The assertion directly follows from Proposition \ref{prop:estbound} and Lemma \ref{lem1}.
\hfill $\Box$

\subsection{Proof of Theorem \ref{theo-smooth1}}

(i) Using Theorem \ref{theo2b}, a Taylor expansion of $g$ of order $\lfloor\beta\rfloor$  and assumption \ref{K1}, one can show by direct calculations that for some $\tau_u\in (0,1)$
\beq
\sup_{x\in I_n}|\tilde g(x)-g(x)|
&\leq&\sup_{x\in I_n}\left|\int_{h_n}^{1-h_n}(\hat g(z)-g(z))\frac 1{b_n}K\left(\frac{x-z}{b_n}\right)dz\right|\\
&&+\sup_{x\in I_n}\left|\int_{h_n}^{1-h_n}(g(z)-g(x))\frac 1{b_n}K\left(\frac{x-z}{b_n}\right)dz\right|\\
&\leq& \sup_{z\in [h_n,1-h_n]}|\hat{g}(z)-g(z)|O(1)+\sup_{x\in I_n}\left|\int_{-1}^{1}(g(x-ub_n)-g(x))K\left(u\right)du\right|\\
&\leq&O(h_n^{\beta})+O_P\Big(\Big(\frac{|\log h_n|}{nh_n}\Big)^{\frac 1\alpha}\Big)\\
&&+b_n^{\lfloor\beta\rfloor}\sup_{x\in I_n}\left|\frac 1{\lfloor\beta\rfloor!}\int_{-1}^1u^{\lfloor\beta\rfloor}(g^{(\lfloor\beta\rfloor)}(x-\tau_u ub_n)-g^{(\lfloor\beta\rfloor)}(x))K(u)du\right|.
\eeq
Now the H\"{o}lder property of $g$ combined by \ref{K1} yields the desired result.

\medskip

(ii) Since $g$ is bounded on $[h_n,1-h_n]$ and $\sup_{x\in[h_n,1-h_n]}|\hat g(x)-g(x)|=o_P(1)$,   $\hat g$ is eventually bounded on $[h_n,1-h_n]$ too. Note that the partial derivative of $\hat g(z)b_n^{-1}K((x-z)/b_n)$ with respect to $x$ is continuous and bounded (for fixed $n$). Thus
we can exchange integration and differentiation and obtain
\[\sup_{x\in I_n}|\tilde{g}'(x)-g'(x)|=\sup_{x\in I_n}\left|\int_{h_n}^{1-h_n}\hat g(z)\frac 1{b_n^2}K'\Big(\frac{x-z}{b_n}\Big)dz-g'(x)\right|.\]
Integration by parts yields
\beq
\int_{h_n}^{1-h_n}g(z)\frac 1{b_n^2}K'\left(\frac{x-z}{b_n}\right)dz
&=&\int_{h_n}^{1-h_n}g'(z)\frac 1{b_n}K\left(\frac{x-z}{b_n}\right)dz
\eeq
since $K(-1)=K(1)=0$. Therefore
\beq
\sup_{x\in I_n}|\tilde{g}'(x)-g'(x)|&\leq&\sup_{x\in I_n}\left|\int_{h_n}^{1-h_n}(\hat g(z)-g(z))\frac 1{b_n^2}K'\left(\frac{x-z}{b_n}\right)dz\right|\\
&&+\sup_{x\in I_n}\left|\int_{h_n}^{1-h_n}(g'(z)-g'(x))\frac 1{b_n}K\left(\frac{x-z}{b_n}\right)dz\right|\\
&\leq& \sup_{z\in [h_n,1-h_n]}|\hat{g}(z)-g(z)|O(b_n^{-1})+\sup_{x\in I_n}\left|\int_{-1}^{1}(g'(x-ub_n)-g'(x))K\left(u\right)du\right|.
\eeq
Similarly as in the proof of (i), assertion (ii) follows by Theorem \ref{theo2b}, a Taylor expansion of $g'$ of order  $\lfloor\beta\rfloor-1$ and the assumptions \ref{K1} and \ref{G1}.

\medskip

(iii) We distinguish the cases $|x-y|>a_n$ and $|x-y|\leq a_n$ for some suitable sequence $(a_n)_{n\in\en}$ with $\lim_{n\to\infty}a_n=0$ specified later. In the first case, we obtain
\begin{eqnarray}
\nonumber
\lefteqn{\sup_{x,y\in I_n,|x- y|>a_n}\frac{|\tilde g'(x)-g'(x)-\tilde g'(y)+g'(y)|}{|x-y|^{\delta}}}\\
\nonumber
&\leq& 2\sup_{x\in I_n}|\tilde{g}'(x)-g'(x)|a_n^{-\delta}\\
&=&\Bigg(O(b_n^{\beta-1})+\Bigg(O(h_n^{\beta})+O_P\left(\left(\frac{|\log h_n|}{nh_n}\right)^{\frac 1\alpha}\right)\Bigg)b_n^{-1}\Bigg)a_n^{-\delta}.
\label{iii-1}
\end{eqnarray}
In the second case, we use a decomposition like in the proof of (ii):
\beq
\lefteqn{\sup_{x,y\in I_n,0<|x- y|\leq a_n}\frac{|\tilde g'(x)-g'(x)-\tilde g'(y)+g'(y)|}{|x-y|^{\delta}}}\\
&\leq&\sup_{x,y\in I_n,0<|x- y|\leq a_n}\frac{\left|\int_{h_n}^{1-h_n}(\hat g(z)-g(z))\frac 1{b_n^2}\left(K'\left(\frac{x-z}{b_n}\right)-K'\left(\frac{y-z}{b_n}\right)\right)dz\right|}{|x-y|^{\delta}}\\
&&+\sup_{x,y\in I_n\atop 0<|x- y|\leq a_n}\frac{|g'(x)-g'(y)|}{|x-y|^{\delta}}
+\sup_{x,y\in I_n\atop 0<|x-y|\leq a_n}\frac{\left|\int_{h_n}^{1-h_n}g'(z)\frac 1{b_n}\left(K\left(\frac{x-z}{b_n}\right)-K\left(\frac{y-z}{b_n}\right)\right)dz\right|}{|x-y|^{\delta}}
.
\eeq
By Lipschitz continuity of $K'$ and Theorem \ref{theo2b},  the first term on the right hand side is of the order
\begin{eqnarray}\label{iii-2}
\Bigg(O(h_n^{\beta})+O_P\Big(\Big(\frac{|\log h_n|}{nh_n}\Big)^{\frac 1\alpha}\Big)\Bigg)\frac 1{b_n^3}O(a_n^{1-\delta}).
\end{eqnarray}
For $\beta\geq 2$, the second term is of the order $a_n^{1-\delta}$ as $g'$ is Lipschitz continuous, while for $\beta\in (1,2)$ assumption \ref{G1} yields the rate
$a_n^{\beta-1-\delta}$. In both cases, condition ({\bf B2.$\boldsymbol\delta$}) ensures that the second term converges to 0.
\\
The last term on the right hand side can be rewritten as
$$\sup_{x,y\in I_n\atop 0<|x-y|\leq a_n}\frac{\left|\int_{-1}^1(g'(x-h_nu)-g'(y-h_nu))K(u)\,du\right|}{|x-y|^{\delta}}$$
and is thus of the same order as the second term by assumption \ref{G1}.


To conclude the proof, one needs to find a sequence $a_n=o(1)$ such that  (\ref{iii-1}) and (\ref{iii-2}) tend to 0 in probability, i.e.
$$ b_n^{\beta-1}+\frac{\vartheta_n}{b_n}=o(a_n^{\delta})\mbox{ and } a_n^{1-\delta}=o\Big(\frac{b_n^3}{\vartheta_n}\Big)$$
with $\vartheta_n:=h_n^\beta+(|\log h_n|/(nh_n))^{1/\alpha}$. Obviously, such a sequence $a_n$ exists if and only if
$$b_n^{\beta-1}+\frac{\vartheta_n}{b_n}=
o\Big(\Big(\frac{b_n^3}{\vartheta_n}\Big)^{\frac{\delta}{1-\delta}}\Big),$$
which in turn is equivalent to condition ({\bf B2.$\boldsymbol\delta$}).
\hfill $\Box$

\subsection{Proof of Theorem \ref{theo3a}}

The assumptions about $\alpha$ ensure that $\beta/(\alpha\beta+1)>1/(2(\alpha\wedge 1))$, and so in view of \eqref{opt-rate} the uniform estimation error of $\hat g$ is stochastically of smaller order than $n^{-1/(2(\alpha\wedge 1))}$. Hence there exists a sequence
\begin{equation}\label{a_n-alpha}
a_n=o(n^{-\frac{1}{2(\alpha\wedge 1)}})
\end{equation}
such that
\[P\Big(\sup_{x\in[h_n,1-h_n]}|\hat g(x)-g(x)|\leq a_n\Big) \nto 1.\]
Let $\bar F_{n}(y,s):=\frac 1{m_n}\sjns I\{\e_j\leq y\}I\{h_n<\frac jn\leq 1-h_n\}$. Since
\beq
\hat{F}_{n}(y,s)
&=&\frac 1{m_n}\sjns I\{\e_j\leq y+(\hat{g}-g)(\textstyle{\frac jn})\}I\{h_n<\textstyle{\frac jn}\leq 1-h_n\}
\eeq
we may conclude
\begin{eqnarray*}
\sqrt n(\bar F_{n}(y-a_n,s)-\bar s_n F_{\ns}(y)) & \leq & \sqrt n(\hat{F}_{n}(y,s)-\bar s_nF_{\ns}(y))\\
 & \leq & \sqrt n(\bar F_{n}(y+a_n,s)-\bar s_nF_{\ns}(y))
\end{eqnarray*}
for all $y\in\er$ and  $s\in[0,1]$ with probability converging to 1.

We take a closer look at the bounds.
The sequential empirical process
\begin{equation} \label{eq:Endef}
    E_n(y,s)=n^{-1/2}\sjns(I\{\e_j\leq y\}-F(y)),\quad y\in\er,s\in[0,1],
\end{equation}
 converges weakly to a Kiefer process; see e.g.\ Theorem 2.12.1 in van der Vaart and Wellner (1996).
 Now, $n\sim m_n$, the asymptotic equicontinuity of the process $E_n$, the H\"{o}lder continuity \ref{F2} and (\ref{a_n-alpha}) imply
\beq
 \lefteqn{\sqrt n\left(\bar F_{n}(y\pm a_n,s)-\bar s_nF_{\ns}(y)\right)}\\
&=&
\frac{n}{m_n}\Big( E_n(y\pm a_n,s\wedge(1-h_n))-E_n(y,s\wedge(1-h_n))-E_n(y\pm a_n,s\wedge h_n)+E_n(y,s\wedge h_n)\Big)\\
&&{}+\sqrt n\bar s_n(F(y\pm a_n)-F(y))+\sqrt n(\bar F_{n}(y,s)-\bar s_nF_{\ns}(y))\\
&=&o_P(1)+\sqrt n(\bar F_{n}(y,s)-\bar s_nF_{\ns}(y))
\eeq
uniformly for all $y\in\er$, $s\in[0,1]$.

 It remains to be shown that
\begin{eqnarray}
 \lefteqn{\sqrt n(\bar F_{n}(y,s)-\bar s_n F_{\ns}(y))} \nn\\
&=&
\frac{\sqrt{n}}{m_n}\sjns (I\{\e_j\leq y\}-F(y))I\{h_n<\frac jn\leq 1-h_n\}-\frac{\sqrt{n}\bar s_n}{\ns}\sjns (I\{\e_j\leq y\}-F(y)) \nn
\\
&=& \Big(\frac{n}{m_n}-1\Big)\big( E_n(y,s\wedge(1-h_n))-E_n(y,s\wedge h_n)\big) \nn \\
&&{}-\Big(\frac{n\bar s_n}{\ns}-1\Big)E_n(y,s) \nn \\
& & { } +\big(E_n(y,s\wedge(1-h_n))- E_n(y,s\wedge h_n)- E_n(y,s)\big)
 \label{eq:threeterms}
\end{eqnarray}
tends to 0 in probability uniformly for all $y\in\er$, $s\in[0,1]$.

The first term vanishes asymptotically, because $E_n$ is uniformly stochastically bounded and $n\sim m_n$.

Next note that $\bar s_n=0$ for $s<h_n$, while for $s\geq h_n$
\begin{equation}  \label{eq:snbarapprox}
  \frac{n\bar s_n}\ns -1 = \frac{\floor{n(s\wedge (1-h_n)}-\floor{nh_n}}{(1-2h_n+O(n^{-1}))\ns} -1 = \frac{O(nh_n)}{(1-2h_n+O(n^{-1}))\ns},
\end{equation}
which is uniformly bounded for all $s\in[h_n,1]$ and tends to 0 uniformly with respect to $s\in[h_n^{1/2},1]$. Moreover, $E_n$ is uniformly stochastically bounded and $\sup_{0\le s\le h_n^{1/2}, y\in\er} |E_n(y,s)|=o_P(1)$, because $E_n$ is asymptotically equicontinuous with $E_n(y,0)=0$. Hence, the second term in \eqref{eq:threeterms} converges to 0 in probability, too. Likewise, the convergence of the last term to 0 follows from the asymptotic equicontinuity of $E_n$, which concludes the proof.
\hfill $\Box$

\subsection{Proof of Theorem \ref{theo-smooth2} and of Remark \ref{rem:withoutB2}}

 For any interval $I\subset\er$ and constant $k>0$, define the following class of differentiable functions:
\[C_k^{1+\delta}(I)=\left\{d:I\to\er\Big|\max\Big\{\sup_{x\in I}|d(x)|,\sup_{x\in I}|d'(x)|,\sup_{x,y\in I,x\neq y}\frac{|d'(x)-d'(y)|}{|x-y|^{\delta}}\Big\}\leq k\right\}.\]
Then Theorem \ref{theo-smooth1} yields $P((\tilde g-g)\in C_{1/2}^{1+\delta}(I_n))\to 1$ as $n\to\infty$. Hence there exist random functions $d_n:[0,1]\to\er$ such that $d_n(x)=(\tilde g-g)(x)$ for all $x\in I_n$ and $P\left(d_n\in C_1^{1+\delta}([0,1])\right)\to 1$ for $n\to\infty$. (For instance, one may extrapolate $\tilde g-g$ linearly on $[0,h_n]$ and on $[1-h_n,1]$.)

On the space $\cf:=\er\times C_1^{1+\delta}([0,1])$ we define the semimetric
\[\rho((y,d),(y^*,d^*))=\max\Big\{\sup_{x\in [0,1]}\sup_{\gamma\in C_1^{1+\delta}([0,1])}\left|F(y+\gamma(x))-F(y^*+\gamma(x))\right|,\sup_{x\in [0,1]}|d(x)-d^*(x)|\Big\}.\]
For $\varphi=(y,d)\in\cf$ let
\[Z_{nj}(\varphi):=\frac{\sqrt n}{m_n}I\{\e_j\leq y+d({\textstyle\frac jn})\}I\{{\textstyle\frac jn}\in I_n\}-\frac1{\sqrt n}I\{\e_j\leq y\}
\]
and
\[G_n(\varphi):=\sjn (Z_{nj}(\varphi)-E[Z_{nj}(\varphi)]).
\]
Note that
\beq
 \lefteqn{G_n(y,d_n)}\\
  & = & \frac{\sqrt{n}}{m_n} \sjn I\{\e_j\leq y-g(j/n)+\tilde g(j/n)\} I\{j/n\in I_n\} \\
 & & {}- \frac{\sqrt{n}}{m_n} \sjn F(y+(\tilde g-g)(j/n)) I\{j/n\in I_n\}- \frac 1{\sqrt{n}}\sjn I\{\e_j\leq y\} + \sqrt{n} F(y) \\
 & = & \sqrt{n}\Big( \tilde F_n(y)- \frac 1n\sjn I\{\eps_j\leq y\} -\frac 1{m_n}\sjn \left(F\left(y+(\tilde g-g)(\textstyle{\frac jn})\right)-F(y)\right)I\{{\textstyle{\frac jn}}\in I_n\}\Big).
\eeq
We will
apply Theorem 2.11.9 of van der Vaart and Wellner (1996) to show that the process $(G_n(\varphi))_{\varphi\in\cf}$ converges to a (Gaussian) limiting process. In particular, $G_n$ is asymptotically equicontinuous, which readily yields the assertion, because $\sup_{y\in\er} \rho((y,d_n),(y,0))=\sup_{x\in[0,1]}|d_n(x)|=o_P(1)$ and the variance of
$$ G_n(y,0) = \frac 1{\sqrt{n}}\sjn \Big(\frac n{m_n}-1\Big)I\{\e_j\le y\}I\{j/n\in I_n\} - I\{\e_j\le y\}I\{j/n\not\in I_n\}
$$
 tends to 0, implying that $G_n(y,0)=o_P(1)$ uniformly in $y$. Thus $G_n(y,d_n)=o_P(1)$ uniformly in $y$ and the assertion holds.

One may proceed as in the proof of Lemma 3 in Neumeyer and Van Keilegom (2009) (see the online supporting information to that article) to prove that the conditions of  Theorem 2.11.9 of van der Vaart and
Wellner (1996) are fulfilled. The proof of the first two displayed formulas of this theorem are analogous. The only difference is that Neumeyer and Van Keilegom (2009) assume a bounded error density while  we use H\"{o}lder continuity of $F$, see assumption  \ref{F2}.
Next we show that the bracketing entropy condition (i.e., the last displayed condition in Theorem 2.11.9 of van der Vaart and
Wellner, 1996) is fulfilled and that $(\cf,\rho)$ is totally bounded.

To this end, let $d_m^L\le d_m^U$, $m=1,\ldots,M$, be brackets for $C_1^{1+\delta}([0,1])$ of length $\eta^{2/(\alpha\wedge 1)}$ w.r.t.\ the supremum norm. According to van der Vaart and
Wellner (1996), Theorem 2.7.1 and Corollary 2.7.2, $M=O\big(\exp(\kappa \eta^{-2/((1+\delta)(\alpha\wedge 1))})\big)$ brackets are needed. For each $m$ define $F_m^L(y):= n^{-1}\sjn F(y+d_m^L(j/n))$ and choose $y_{m,k}^L$, $k=1,\ldots,K=O(\eta^{-2})$ such that $F_m^L(y_{m,k}^L)- F_m^L(y_{m,k-1}^L)<\eta^2$ for all $k\in\{1,\ldots,K+1\}$ with $y_{m,0}^L:=-\infty$ and $y_{m,K+1}^L:=\infty$. Define $F_m^U$ and $y_{m,k}^U$ analogously, $\tilde y_{m,k}^L:= y_{m,k}^L$ and denote by $\tilde y_{m,k}^U$ the smallest $y_{m,l}^U$  larger than or equal to $y_{m,k+1}^L$. Then $\cf$ is covered by
$$ \cf_{mk} := \{(y,d)\in\cf \mid \tilde y_{m,k}^L\le y\le \tilde y_{m,k}^U, d_m^L\le d\le d_m^U\}, \quad m=1,\ldots,M,\; k=1,\ldots,K.
$$
Check that by condition \ref{F2}
\begin{eqnarray}
 \sup_{y\in\er} |F_m^U(y)-F_m^L(y)| & \le & \sup_{y\in\er} n^{-1}\sjn |F(y+d_m^U(j/n))- F(y+d_m^L(j/n))| \nonumber\\
 & \le & L_F \sup_{x\in\er} |d_m^U(x)-d_m^L(x)|^{\alpha\wedge 1} \le L_F\eta^2 \label{eq:functdiff}
\end{eqnarray}
with $L_F$ denoting the H\"{o}lder constant of $F$. Thus
\beq
  \lefteqn{\frac 1n \sjn E\Big[\sup_{(y,d),(y^*,d^*)\in\cf_{mk}}\big|I\{\e_j\le y+d(j/n)\}-I\{\e_j\le y^*+d^*(j/n)\}\big|\Big]^2} \\
  & \le & \frac 1n \sjn E\big[I\{\e_j\le \tilde y_{m,k}^U+d_m^U(j/n)\}-I\{\e_j\le \tilde y_{m,k}^L+d_m^L(j/n)\}\big]^2 \\
  & \le & F_m^U(\tilde y_{m,k}^U)-F_m^L(\tilde y_{m,k}^L) \\
  & \le & |F_m^U(\tilde y_{m,k}^U)-F_m^U(\tilde y_{m,k+1}^L)| +
  |F_m^U(\tilde y_{m,k+1}^L)-F_m^L(\tilde y_{m,k+1}^L)| +
  |F_m^L(\tilde y_{m,k+1}^L)-F_m^L(\tilde y_{m,k}^L)|\\
  & \le & (2+L_F)\eta^2
\eeq
where the last step follows from \eqref{eq:functdiff} and the definitions of $\tilde y_{m,k}^L$ and $\tilde y_{m,k}^U$. Hence we obtain for the squared diameter of $\cf_{mk}$ w.r.t.\ $L_2^n$
\beq
   \lefteqn{\sjn E\Big[\sup_{(y,d),(y^*,d^*)\in\cf_{mk}}|Z_{nj}(y,d)-Z_{nj}(y^*,d^*)|\Big]^2} \\
   & \le & 2 \frac n{m_n^2} \sjn E\Big[\sup_{(y,d),(y^*,d^*)\in\cf_{mk}}\big|I\{\e_j\le y+d(j/n)\}-I\{\e_j\le y^*+d^*(j/n)\}\big|\Big]^2 I\{j/n\in I_n\} \\
   & & {} + \frac 2n \sjn E\Big[\sup_{(y,d),(y^*,d^*)\in\cf_{mk}}\big|I\{\e_j\le y\}-I\{\e_j\le y^*\}\big|\Big]^2 \\
   & \le & 3(2+L_F)\eta^2
\eeq
for sufficiently large $n$.
This shows that the bracketing number satisfies $\log N_{[\,]}(\eta,\cf,L_2^n)=O(\log M +\log K)=O\big(\eta^{-2/((1+\delta)(\alpha\wedge 1))}\big)$, and the last displayed condition of Theorem 2.11.9 of van der Vaart and Wellner (1996) follows from $\delta>1/\alpha-1$.

It remains to show that $(\cf,\rho)$ is totally bounded,
i.e.\ that, for all $\eta\in (0,1)$, the space $\cf$ can be covered by finitely many sets with $\rho$-diameter less than $5\eta$. To this end, choose $d_m^L$ and $d_m^U$ as above. For each $m\in\{1,\ldots,M\}$ and $j\in\{0,\ldots,J:=\ceil{\eta^{-1}}\}$, let $s_j:=j\eta^{1/(\alpha\wedge 1)}\wedge 1$ and $F_{jm}(y):=P(\e_1\le y+d_m^L(s_j))$,  and choose an increasing sequence $y_{jm,k}$, $k=1,\ldots,K:=\floor{\eta^{-1}}$, such that
$F_{jm}(y_{jm,k})- F_{jm}(y_{jm,k-1})<\eta$ for all $k\in\{1,\ldots,K+1\}$ with $y_{jm,0}:=-\infty$ and $y_{jm,K+1}:=\infty$. Denote by $\bar y_l$, $1\le l\le L$, all points $y_{jm,k}$,  $j\in\{0,\ldots,J\}$, $m\in\{1,\ldots,M\}$, $k\in\{1,\ldots,K\}$, in increasing order. We show that all sets
$\cf_{ml}:=\{ (y,d)\mid \bar y_{l-1}\le y\le \bar y_l, d_m^L\le d\le d_m^U\}$ have $\rho$-diameter less than $5\eta$.
Check that, for all $1\le l\le L$, one has
\begin{eqnarray*}
 \lefteqn{\sup_{x\in[0,1]} \sup_{\gamma\in C_1^{1+\delta}([0,1])} |F(\bar y_l+\gamma(x))-F(\bar y_{l-1}+\gamma(x))|} \nn\\
 & \le & \max_{1\le j\le J} \sup_{s_{j-1}\le x\le s_j} \max_{1\le m\le M} \sup_{d_m^L\le \gamma\le d_m^U} \Big[ |F(\bar y_l+\gamma(x))-F(\bar y_l+\gamma(s_j))| \nn \\
 & & + |F(\bar y_l+\gamma(s_j))-F(\bar y_l+d_m^L(s_j))| + |F(\bar y_l+d_m^L(s_j))-F(\bar y_{l-1}+d_m^L(s_j))| \nn\\
 & & + |F(\bar y_{l-1}+d_m^L(s_j))-F(\bar y_{l-1}+\gamma(s_j))| + |F(\bar y_{l-1}+\gamma(s_j))-F(\bar y_{l-1}+\gamma(x))|\Big] \nn \\
 & < & \max_{1\le j\le J} \Big[ (s_j-s_{j-1})^{\alpha\wedge 1} + \eta^2 + \eta + \eta^2 + (s_j-s_{j-1})^{\alpha\wedge 1}\Big] \label{eq:totbdd}\\
 & \le & 5\eta. \nn
\end{eqnarray*}
Therefore, for all $(y,d), (y^*,d^*)\in\cf_{ml}$
\beq
 \lefteqn{\rho((y,d), (y^*,d^*))}\\
  & \le & \max\Big\{\sup_{x\in[0,1]} \sup_{\gamma\in C_1^{1+\delta}([0,1])} |F(\bar y_l+\gamma(x))-F(\bar y_{l-1}+\gamma(x))|,\sup_{x\in[0,1]} d_m^U(x)-d_m^L(x)\Big\} \\
 & \le & \max\{5\eta,\eta^{2/(\alpha\wedge 1)}\}= 5\eta,
\eeq
 which concludes the proof of Theorem \ref{theo-smooth2}.

 If we drop the assumption $\delta>1/\alpha-1$ but require $F$ to be Lipschitz continuous, then we use brackets for $C_1^{1+\delta}([0,1])$ of length $\eta^2$ (instead of $\eta^{2/(\alpha\wedge 1)}$) and replace \eqref{eq:functdiff} with
 \begin{eqnarray*}
 \sup_{y\in\er} |F_m^U(y)-F_m^L(y)| & \le & \sup_{y\in\er} n^{-1}\sjn |F(y+d_m^U(j/n))- F(y+d_m^L(j/n))| \nonumber\\
 & \le & L_F \sup_{x\in\er} |d_m^U(x)-d_m^L(x)| \le L_F\eta^2
\end{eqnarray*}
with $L_F$ denoting the Lipschitz constant of $F$ to prove $\log N_{[\,]}(\eta,\cf,L_2^n)=O\big(\eta^{-2/(1+\delta)}\big)$, which again yields the third condition of Theorem  2.11.9 of van der Vaart and Wellner (1996). Likewise, in the last part of the proof, one defines $s_j:=j\eta\wedge 1$ and replaces \eqref{eq:totbdd} with $\max_{1\le j\le J} \big[ (s_j-s_{j-1}) + \eta^2 + \eta + \eta^2 + (s_j-s_{j-1})\big]\le 5\eta$.
 \hspace*{\fill}$\Box$

\bigskip

In the remaining proofs to Section \ref{section3}, we use the index $n$ for the estimators to emphasis the dependence on the sample size and to distinguish between estimators and polynomials corresponding to a given sample on the one hand and corresponding objects in a limiting setting on the other hand.

\subsection{Proof of Lemma \ref{lem:expectg0}}

 Proposition \ref{prop:estbound} and the proof of Lemma \ref{lem1} show that there exist constants $d,\tilde d>0$ depending only on $\beta$ and $c_g$ such that $E(\hat g_n(x))\le \tilde d E(M_{n,0})$ and $P\{M_{n,0}>t\}\le \big(1+dnh_n(1-F(-t))\big)(F(-t))^{dnh_n}$ for all $t>0$.

 Let $a_n:= a(\log n/(nh_n))^{1/\alpha}$ for a suitable constant $a>0$ and fix some $t_0>0$ such that $(1-F(-t))/(ct^\alpha)\in(1/2,2)$ for all $t\in(0,t_0]$. Then
 \begin{eqnarray*}
  \lefteqn{E(M_{n,0})}\\
   & = & \int_0^\infty P\{M_{n,0}>t\}\, dt \\
  & \le & a_n+ \int_{a_n}^{t_0} \big(1+dnh_n(1-F(-t))\big)(F(-t))^{dnh_n}\, dt + (1+dnh_n)\int_{t_0}^\infty((F(-t))^{dnh_n}\, dt.
  \end{eqnarray*}
  Now, for sufficiently large $n$,
  \begin{eqnarray*}
     \lefteqn{\int_{a_n}^{t_0} \big(1+dnh_n(1-F(-t))\big)(F(-t))^{dnh_n}\, dt }\\
     & \le & \int_{a_n}^{t_0} \big(1+2cdnh_nt^\alpha\big)\Big(1-\frac{c}2 t^\alpha\Big)^{dnh_n}\, dt\\
     & \le & (1+2cd) nh_n \int_{a_n}^{t_0} t^\alpha  \exp\Big(-\frac{c}2 dnh_nt^\alpha\Big)\, dt\\
     & \le & (1+2cd) nh_n \frac{t_0}\alpha \int_{a_n^\alpha}^{t_0^\alpha}\exp\Big(-\frac{c}2 dnh_nu\Big)\, du\\
     & \le & \frac{2(1+2cd) nh_n t_0}{\alpha c d n h_n} \exp\Big(-\frac{c}2 da^\alpha \log n\Big)\\
     &= & o(n^{-\xi})
  \end{eqnarray*}
  for all $\xi>0$ if $a$ is chosen sufficiently large. Hence the assertion follows from
  \ref{H1} and \ref{F3} which imply
  $$ \int_{t_0}^\infty (F(-t))^{dnh_n}\, dt\le nh_n((F(-t_0))^{dnh_n}+\int_{nh_n}^\infty t^{-d\tau nh_n}\, dt = o(n^{-\xi})
  $$
  for all $\xi>0$. \hfill$\Box$

\subsection{Proof of Theorem \ref{theo:remainderterm}}

As the density $f$ is bounded and Lipschitz continuous, one has
\begin{eqnarray*}
 \big|F\big(y+(\gts-g)(j/n))-F(y)-f(y)(\gts-g)(j/n)\big| & = & \Big|\int_0^{(\gts-g)(j/n)} f(y+t)-f(y)\, dt\Big|\\
  & = & O\big( (\gts-g)^2(j/n)\big)
\end{eqnarray*}
uniformly for $y\le y_0$ and $j/n\in I_n$. Hence the remainder term can be approximated by a sum of estimation errors as follows:
\begin{eqnarray*}
  \lefteqn{\bigg| \frac 1{m_n} \sum_{j=1}^n \big(F\big(y+(\gts-g)(j/n))-F(y)\big)I\{j/n\in I_n\} -
  \frac{f(y)}{m_n} \sum_{j=1}^n (\gts-g)(j/n)I\{j/n\in I_n\} \bigg|}\\
  & = & O\bigg( \frac 1{m_n} \sum_{j=1}^n (\gts-g)^2(j/n)I\{j/n\in I_n\} = O_P\Big(h_n^{2\beta}+b_n^{2\beta}+\Big(\frac{\log n}{nh_n}\Big)^{2/\alpha}\Big) = o_P(n^{-1/2}) \hspace*{2cm}
\end{eqnarray*}
where for the last conclusions we have used Theorem \ref{theo2b}, Lemma \ref{lem:expectg0} and the assumptions \ref{H2} and \ref{B3}.
Thus the assertion follows if we show that
$$\frac 1{m_n} \sum_{j=1}^n (\gts-g)(j/n)I\{j/n\in I_n\} = o_P(n^{-1/2}). $$

To this end, note that $\gts(x)$ and $\gts(y)$ are independent for $|x-y|>2(h_n+b_n)$. For simplicity, we assume that $2n(h_n+b_n)=:k_n$ is a natural number. If we split the whole sum into blocks with $k_n$ consecutive summands, then all blocks with odd numbers are independent and all blocks with even numbers are independent. It suffices to show that
\begin{eqnarray*}
  \frac 1{m_n}\sum_{\ell=1}^{\floor{n/(2k_n)}} \Delta_{n,2\ell-1} & =  & o_P(n^{-1/2})\\
  \frac 1{m_n}\sum_{\ell=1}^{\floor{n/(2k_n)}} \Delta_{n,2\ell} & =  & o_P(n^{-1/2})
\end{eqnarray*}
where $\Delta_{n,l}=\sum_{j=(l+1)k_n}^{(l+2)k_n-1} (\gts-g)(j/n)$, $1\le \ell\le\floor{n/k_n}$. We only  consider the second sum, because the first convergence obviously follows by the same arguments.

It suffices to verify
\begin{eqnarray}
  E\big(\Delta_{n,2\ell}^2\big) &=& o(k_n) \label{eq:wlln_cond1}\\
  E\big(\Delta_{n,2\ell}\big) & = & o\big(n^{-1/2}k_n\big)=o\big(n^{1/2}(h_n+b_n)\big) \label{eq:wlln_cond2}
\end{eqnarray}
uniformly for all $1\le \ell\le \floor{n/(2k_n)}$, since then
$$ E\Big(\sum_{\ell=1}^{\floor{n/(2k_n)}}\Delta_{n,2\ell} \Big)^2 = \sum_{\ell=1}^{\floor{n/(2k_n)}} Var(\Delta_{n,2\ell})+ \Big(\sum_{\ell=1}^{\floor{n/(2k_n)}} E(\Delta_{n,2\ell})\Big)^2 = o(n),
$$
which implies the assertion.

To prove \eqref{eq:wlln_cond1}, note that according to Lemma \ref{lem:expectg0}, Proposition \ref{prop:estbound} and the proofs of Lemma \ref{lem1} and  of Theorem \ref{theo-smooth1}(i), there exist constants $c_1,c_2,c_3>0$ (depending only on $\beta$, $c_g$ and the kernel $K$) such that
 $$\sup_{x\in I_n} |\gts(x)-g(x)|\le c_1\Big(h_n^\beta+b_n^\beta+\Big(\frac{\log n}{nh_n}\Big)^{1/\alpha}+\max(M_1^*,M_2^*)\big)
 $$
 where $M_1^*,M_2^*$ are independent random variables such that $P\{M_i^*>t\}\le 1-(1-P\{M_{n,0}>t\})^{c_2(h_n+b_n)/h_n}$ with
 $$ P\{M_{n,0}>t\}\le \big(1+c_3nh_n(1-F(-t))\big)(F(-t))^{c_3nh_n}.
 $$
 Because $k_n\big(h_n^\beta+b_n^\beta+(\log n/(nh_n))^{1/\alpha}\big)=o(k_n^{1/2})$ by \ref{H2} and \ref{B3}, it suffices to show that
   \begin{equation} \label{eq:wlln_cond1b}
    E\big((M_i^*)^2\big) = \int_0^\infty P\{M_i^*>t^{1/2}\}\, dt = o(1/k_n).
  \end{equation}

  Fix some $t_0\in(0, (2c)^{-2/\alpha})$ such that $(1-F(-t))/(ct^\alpha)\in(1/2,2)$ for all $t\in(0,t_0]$. In what follows, $d$ denotes a generic constant (depending only on $\beta,c_g,c$ and $K$) which may vary from line to line. Applying the inequalities $\exp(-2\rho u)\le (1-u)^\rho\le\exp(-\rho u)$, which holds for all $\rho>0$ and $u\in (0,1/2)$,  we obtain for $(nh_n/\log n)^{-2/\alpha}<t\le t_0$ and sufficiently large $n$
  \begin{eqnarray*}
    P\{M_i^*>t^{1/2}\} & \le & 1-\big[1-(1+c_3nh_n 2ct^{\alpha/2})(1-c t^{\alpha/2}/2)^{c_3nh_n}\big]^{c_2(h_n+b_n)/h_n}\\
    & \le & 1-\big[1-3c_3c nh_n t^{\alpha/2}\exp\big(-c_3cnh_nt^{\alpha/2}/2\big)\big]\\
    & \le & 1-\exp\Big(-d n(h_n+b_n)t^{\alpha/2}\exp\big(-c_3cnh_nt^{\alpha/2}/2\big)\Big)\\
    & \le & d n(h_n+b_n)t^{\alpha/2}\exp\big(-c_3cnh_nt^{\alpha/2}/2\big).
  \end{eqnarray*}
  Therefore, for sufficiently large $a>0$,
  \begin{eqnarray}
    \lefteqn{\int_0^{t_0^2} P\{M_i^*>t^{1/2}\}\, dt} \nonumber\\
    & \le & a\Big(\frac{nh_n}{\log n}\Big)^{-2/\alpha} + dt_0 n(h_n +b_n) \int_{a(nh_n/\log n)^{-2/\alpha}}^{t_0} t^{\alpha/2-1}\exp\big(-c_3cnh_nt^{\alpha/2}/2\big)\, dt \nonumber\\
    & \le & o(1/(n(h_n+b_n)))+dt_0 n(h_n +b_n) \exp\big(-c_3ca^{\alpha/2}\log n/2\big) \nonumber\\
    & = & o(1/(n(h_n+b_n)))  \label{eq:Misintbd}
  \end{eqnarray}
  where in the last but one step we apply the conditions \ref{B3} and \ref{H2}.
  Now, assertion \eqref{eq:wlln_cond1b} (and hence \eqref{eq:wlln_cond1}) follows from
  \begin{eqnarray*}
    \int_{t_0^2}^\infty P\{M_i^*>t^{1/2}\}\, dt
    & \le & \int_{t_0^2}^\infty  1-\Big[1-c_3nh_n(F(-t^{1/2}))^{c_3nh_n}\Big]^{c_2(h_n+b_n)/h_n}\, dt\\
    & \le & \int_{t_0^2}^\infty 1-\exp\Big(-dn(h_n+b_n) (F(-t^{1/2}))^{c_3nh_n}\Big)\, dt\\
    & \le & d n(h_n+b_n) \Big(nh_n(F(-t_0))^{c_3nh_n}+\int_{nh_n}^\infty t^{-\tau c_3 nh_n/2} \, dt\Big)\\
    & = & o(n^{-\xi})
  \end{eqnarray*}
  for all $\xi>0$ and sufficiently large $n$, where we have used \ref{H2} and \ref{F3}.

  To establish \eqref{eq:wlln_cond2}, first note that for a kernel $K$ of order $d+1$ with $d:=\floor{\beta}$
  \begin{eqnarray*}
   E(\tilde g_n(x)-g(x)) & = &
    E\Big( \int_{-1}^1 \Big(\hat g_n(x+b_nu)-\sum_{j=0}^d \frac{g^{(j)}(x)}{j!} (b_nu)^j\Big) K(u)\, du \\
    & = &  \int_{-1}^1 E(\hat g_n(x+b_nu)-g(x+b_nu))K(u)\, du + O(b_n^\beta)
  \end{eqnarray*}
  uniformly for all $x\in[h_n+b_n,1-h_n-b_n]$.
  In view of \ref{K1}, \ref{H2} and \ref{B3}, it thus suffices to show that
  \begin{equation} \label{eq:wlln_cond2b}
   \big|E(\hat g_n(x)-g(x))-E_{g\equiv 0}(\hat g_n(1/2))\big| =  \big|E(\hat g_n(x)-g(x))-E_{g\equiv 0}(\hat g_n(x))\big|= o(n^{-1/2})
  \end{equation}
  uniformly for Lebesgue almost all $x\in[h_n,1-h_n]$. Note that the distribution of $\hat g_n(x)$ does not depend on $x$ if $g$ equals 0.

  Recall that $\hat g_n(x)=\tilde p_n(0)$ where $\tilde p_n$ is a polynomial on $[-1,1]$  of degree $d$ that solves the linear optimization problem
  $$ \int_{-1}^1 \tilde p_n(t)\, dt \to \min! $$
  under the constraints
  $$ \tilde p_n\Big(\frac{i/n-x}{h_n}\Big) \geq Y_i, \quad \forall\, i\in [n(x-h_n),n(x+h_n)].
  $$
  Define  polynomials
  $$
   q_x(t)  :=   \sum_{k=0}^d \frac 1{k!} g^{(k)}(x)(h_nt)^k, \quad p_n(t):=(nh_n)^{1/\alpha}(\tilde p_n(t)-q_x(t)),\quad t\in[-1,1].
  $$
  Then $q_x((u-x)/h_n)$ is the Taylor expansion of order $d$ of $g(u)$ at $x$ and the estimation error can be written as
  \begin{equation}  \label{eq:esterrorpol}
   \hat g_n(x)-g(x) = (nh_n)^{-1/\alpha}p_n(0).
  \end{equation}
  Note that $p_n$ is a polynomial of degree $d$ that solves the linear optimization problem
  $$ \int_{-1}^1 p_n(t)\,dt \to \min! $$
  subject to
  \begin{equation}  \label{eq:constraint1}
    p_n\Big(\frac{i/n-x}{h_n}\Big)\geq (nh_n)^{1/\alpha} \bar\eps_i,\quad \forall\, i\in [n(x-h_n),n(x+h_n)],
  \end{equation}
  with
  $$ \bar \eps_i := \eps_i+g(i/n)-q_x\Big(\frac{i/n-x}{h_n}\Big). $$
  We now use  point process techniques to analyze the asymptotic behavior of this linear program.

  Denote by
  $$ N_n := \sum_{i\in[n(x-h_n),n(x+h_n)]} \delta_{\textstyle ((i/n-x)/h_n, (nh_n)^{1/\alpha}\bar\eps_i)}
  $$
  a point process of standardized error random variables. Then the constraints \eqref{eq:constraint1} can be reformulated as $N_n(A_{p_n})=0$ where $A_f:=\{(t,u)\in[-1,1]\times\R\mid u>f(t)\}$ denotes the open epigraph of a function $f$.

  Since by \ref{H2}
  $|\bar\eps_i-\eps_i|= g(i_n)-q_x((i/n-x)/h_n))=O(h_n^\beta)=o((nh_n)^{-1/\alpha})$ uniformly for all $i\in[n(x-h_n),n(x+h_n)]$, one has
  $$ E\big(N_n([-1,1]\times(-1,\infty)))\sim 2nh_n P\big\{\bar\eps_1>-(nh_n)^{-1/\alpha}\big\} \to 2c.
  $$
  Therefore, $N_n$
  converges weakly to a Poisson process $N$ on $[-1,1]\times\R$ with intensity measure  $ 2cU_{[-1,1]}\otimes \nu_\alpha$ where $\nu_\alpha$ has Lebesgue density $x\mapsto \alpha |x|^{\alpha-1}I(-\infty,0)$  (see, e.g., Resnick (2007), Theorem 6.3).
  By Skorohod's representation theorem, we may assume that the convergence holds a.s.

  Next we analyze the corresponding linear program in the limiting model to minimize $\int_{-1}^1 p(t)\, dt$ over polynomials of degree $d$ subject to $N(A_p)=0$. In what follows we use a representation of the Poisson process as $N=\sum_{i=1}^\infty \delta_{(T_i,Z_i)}$ where $T_i$ are independent random variables which are uniformly distributed on $[-1,1]$.

  First we prove  by contradiction that the optimal solution is almost surely unique.
  Suppose that there exist more than one solution. From the theory of linear programs it is known that then there exists a solution $p$ such that $J:=\{j\in\N\mid p(T_j)=Z_j\}$ has at most $d$ elements. Because $p$ is bounded and $N$ has a.s.\ finitely many points in any bounded set, $\eta:=\inf\{|p(T_i)-Z_i|\mid i\in\N\setminus J\}>0$ a.s. Since $p$ is an optimal solution, all polynomials $\Delta$ of degree $d$ such that $\Delta(T_j)=0$, $j\in J$, and $\|\Delta\|_\infty<\eta$ must satisfy $\int_{-1}^1\Delta(t)\, dt=0$, because both $p+\Delta$ and $p-\Delta$ satisfy the constraints $N(A_{p\pm\Delta})=0$. In particular, for all polynomials $q$ of degree $d-|J|$, $\Delta(t)=\tau\prod_{i\in J}(t-T_i)q(t)$ is of that type if $\tau>0$ is sufficiently small. Write $\prod_{i\in J}(t-T_i)$ in the form $t^{|J|}+\sum_{l=0}^{|J|-1}a_lt^l$. Then necessarily
  $$ \int_{-1}^1 \prod_{i\in J}(t-T_i)t^j\, dt = \frac2{|J|+j+1}I\{|J|+j\text{ even}\}+ \sum_{l=0}^{|J|-1} \frac{2a_l}{l+j+1} I\{l+j \text{ even}\}=0,
  $$
  for all $j\in\{0,\ldots d-|J|\}$.
  This implies that $(T_i)_{i\in J}$ lies on a  manifold $M_{|J|,d}$ of dimension $|J|-(d-|J|+1)=2|J|-d-1$ which only depends on $|J|$ and $d$. However, by Proposition \ref{prop:estbound},  $\|p\|_\infty \le K_d Z_{\max}$ where
  $$ Z_{\max} := \max_{1\le i\le j_d} \min\{ |Z_i|\mid T_i\in[-1+(j-1)/j_d,-1+j/j_d]\}.
  $$
   The above conclusion contradicts
  $ P\{Z_{\max}>K\}\to 0$ as $K\to\infty$, since
  $$ P\big\{\exists J\subset \N: |J|\le d, (T_j)_{j\in J}\in M_{|J|,d}, \max_{j\in J} |Z_j|\le K_d K\big\}=0
  $$
   for all $K>0$
  (i.e., the fact that among finitely many values $T_i$ a.s.\ there does not exist a subset which lies on a given manifold of lower dimension).

  Therefore the solution $p$ must be a.s.\ unique which in turn implies that it is a basic feasible solution, i.e.,  $|J|\geq d+1$. On the other hand, because the intensity measure of $N$ is absolutely continuous, $|J|\le d+1$ a.s.\ and thus $|J|=d+1$.
  Because of $N_n\to N$ a.s.,  one has $N_n([-1,1]\times[-K_dZ_{\max},\infty))= N([-1,1]\times[-K_dZ_{\max},\infty))=:M$ for sufficiently large $n$. Moreover, one can find a numeration of the points $(T_{n,i},Z_{n,i})$, $1\le i\le M$, of $N_n$ and $(T_{i},Z_{i})$, $1\le i\le M$, of $N$ in $[-1,1]\times[-K_dZ_{\max},\infty)$ such that $(T_{n,i},Z_{n,i})\to (T_i,Z_i)$.

  Next we prove that the solution to the linear program to minimize $\int_{-1}^1 p_n(t)\, dt $ subject to $N_n(A_{p_n})=0$ is eventually unique with $p_n\to p$ a.s.
  Since any optimal solution can be written as a convex combination of  basic feasible solutions, w.l.o.g.\ we may assume that $J_n:=\{1\le i\le M\mid p_n(T_{n,i})=Z_{n,i}\}$ has at least $d+1$ elements. The polynomial $p_n$ is uniquely determined by this set $J_n$. Suppose that along a subsequence $n'$ the set $J_{n'}$ is constant, but not equal to $J$. Then $p_n'$ converges uniformly to the polynomial $\bar p$ of degree $d$ that is uniquely determined by the conditions $\bar p(T_i)=Z_i$ for all $i\in J_{n'}$. In particular, $\bar p$ is different from the unique optimal polynomial $p$ for the limit Poisson process, but it satisfies the constraints $N(A_p)=0$. Thus $\int_{-1}^1 \bar p(t)\, dt>\int_{-1}^1 p(t)\, dt$. On the other hand, for all $\eta>0$ the polynomial $p+\eta$ eventually satisfies the constraints $N_n(A_{p+\eta})= 0$ and thus  $\int_{-1}^1 p(t)+\eta\, dt\geq \int_{-1}^1 \bar p_n(t)\, dt$, which leads to a contradiction.

  Hence, $J_n=J$ for all sufficiently large $n$ and the optimal solution $p_n$ for $N_n$ is unique and it converges uniformly to the optimal solution $p$ for the Poisson process $N$. Moreover, using the relation $(p_n(T_{n,j}))_{j\in J}=(Z_{n,j})_{j\in J}$ (which is a system of linear equation in the coefficients of $p_n$), $p_n(0)$ can be calculated as $w_n^t (Z_{n,j})_{j\in J}$ for some vector $w_n$ which converges to a limit vector $w$ (corresponding to the analogous relation for $p$).

   Exactly the same arguments apply if we replace $\bar\eps_i$ with $\eps_i$, which corresponds to the case that $g$ is identical 0. Since the points $(\tilde T_{n,i},\tilde Z_{n,i})$ of the pertaining point process equal $\big(T_{n,i}, Z_{n,i}-(nh_n)^{1/\alpha}(g(i/n)-q_x((i/n)-x)/h_n)\big)$ and thus $|\tilde Z_{n,i}-Z_{n,i}|\le c_g(nh_n)^{1/\alpha}h_n^\beta$, the difference of the resulting values for optimal polynomial at 0 is bounded by a multiple of $(nh_n)^{1/\alpha}h_n^\beta$. In view of \eqref{eq:esterrorpol} and \ref{H2},  we may conclude that the difference between the estimation errors can be bounded by a multiple of $h_n^\beta=o(n^{-1/2})$, which finally yields \eqref{eq:wlln_cond2b} and thus the assertion.
\hfill$\Box$

\subsection{Proof of Corollary \ref{cor-indep}}

Note that $\textstyle{\frac{\ns}{\sqrt{n}}}(F_{\ns}(y)-F_n(y))=E_n(y,s)-\textstyle{\frac{\ns}n}E_n(y,1)$ with $E_n$ defined in \eqref{eq:Endef}. A similar reasoning as in the proof of Theorem \ref{theo3a} (see \eqref{eq:snbarapprox}) shows that
\begin{equation*}
\sup_{y\in\er,s\in [0,1]}\Big|\Big({{\frac{n\bar s_n}{\ns}}}-1\Big){\frac{\ns}{\sqrt n}}(F_{\ns}(y)-F_n(y))\Big|=o_P(1).
\end{equation*}
Hence, by Theorem \ref{theo3a},  uniformly for all $y\in\er$, $s\in[0,1]$,
\beq
  \lefteqn{\sqrt{n}(\hat F_n(y,s)-\bar s_n\hat F_n(y))}\\
  &=&\sqrt{n}(\hat F_n(y,s)-\bar s_nF_{\ns}(y))-\bar s_n\sqrt{n}((\hat F_n(y)-F_n(y)))+\bar s_n\sqrt{n}(F_{\ns}(y)-F_n(y))\\
&=&\textstyle{\frac{n\bar s_n}{\ns}\frac{\ns}{\sqrt{n}}}(F_{\ns}(y)-F_n(y))+o_P(1)\\
&=& E_n(y,s)-\textstyle{\frac{\ns}{n}}E_n(y,1)+o_P(1)\\
&=& E_n(y,s)-sE_n(y,1)+o_P(1)
\eeq
which converges weakly to $K_F(y,s)-sK_F(y,1)$ for the Kiefer process $K_F$ defined in Theorem \ref{theo3a}. Check that this Gaussian process has the same law as $G(s,F(y))$, because they have the same covariance function.
Thus the Kolmogorov-Smirnov statistic $T_n$ converges weakly to $\sup_{s\in[0,1],y\in\er}|G(s,F(y)|=\sup_{s\in[0,1],z\in[0,1]}|G(s,z)|$,
where the last equality holds by the continuity of $F$.
\hfill $\Box$

\subsection{Proof of Theorem \ref{theo-mon}}

Note that under the given assumptions, the statements of Theorem \ref{theo-smooth1} (i) and (ii) are valid  with rate $o_P(1)$. Let $\Omega_n:=\{\inf_{x\in I_n}\tilde g'(x)>0\}$. From assumption \ref{I1} and  Theorem \ref{theo-smooth1} (ii) it follows that $P(\Omega_n)\to 1$ for $n\to\infty$. But on $\Omega_n$ the estimators $\tilde g_I$ and $\tilde g$ are identical, and thus $\tilde F_{I, \lfloor ns\rfloor}=\tilde F_{\lfloor ns\rfloor}$. Now \eqref{eq:tildeFconv} can be concluded as in the proof of Theorem \ref{theo3a}, because Theorem \ref{theo-smooth1} (i) yields $\sup_{x\in I_n}|\tilde g(x)-g(x)|=o_P(n^{-1/(2(\alpha \wedge 1))})$.
The convergence of the Kolmogorov-Smirnov test statistic then follows exactly as in the proof of Corollary \ref{cor-indep}.
\hfill$\Box$

\end{appendix}

\section*{Acknowledgement}
Financial support by the DFG (Research Unit FOR 1735 {\it Structural Inference in
Statistics: Adaptation and Effciency}) is gratefully acknowledged.

\section*{References}

\begin{description}
\item Akritas, M. and Van Keilegom, I. (2001). Nonparametric estimation of the residual distribution. \textit{Scand.\ J.\ Statist.} {\bf 28}, 549--567.

\item  Anevski, D. and Foug\`{e}res, A.-L.  (2007). Limit properties of the monotone rearrangement for density and regression function estimation. arXiv:0710.4617v1

\item Birke, M. and Neumeyer, N. (2013). Testing Monotonicity of Regression Functions - An Empirical Process Approach. \textit{Scand.\ J.\ Statist.} {\bf 40}, 438--454.

\item Birke, M., Neumeyer, N. and Volgushev, S. (2016+). The independence process in conditional quantile location-scale models and an
application to testing for monotonicity. Statistica Sinica, to appear.

\item Chernozhukov, V., Fern\'{a}ndez-Val, I. and Galichon, A. (2009).
Improving point and interval estimators of monotone functions by rearrangement. \textit{Biometrika} {\bf 96}, 559--575.


\item Daouia, A., Noh, H. and Park, B. U. (2016). Data envelope fitting with constrained polynomial splines. \textit{J. R. Stat. Soc. B.} 78, 3--30.

\item Einmahl, J.\,H.\,J. and Van Keilegom, I. (2008). Specification tests in nonparametric regression. {\em Journal of Econometrics} {\bf 143}, 88--102.

\item F\"{a}re, R. and Grosskopf, S. (1983). Measuring output efficiency. {\em European Journal of Operational Research} {\bf 13}, 173--179.

\item  Gijbels, I. (2005). Monotone regression. In: N. Balakrishnan, S. Kotz, C.B. Read and B. Vadakovic (eds), {\em The Encyclopedia of Statistical Sciences}, 2nd edition. Hoboken, NJ: Wiley.

\item Gijbels, I., Mammen, E., Park, B.\ and Simar, L. (2000). On
estimation of monotone and concave frontier functions. \textit{J.\ Amer.\ Statist.\ Assoc.} {\bf 94}, 220--228.

\item Gijbels, I.\ and Peng, L.  (2000). Estimation of a support curve via order statistics. \textit{Extremes} {\bf 3}, 251--277.

\item Girard, S.\ and Jacob, P. (2008). Frontier estimation via kernel regression on high
power-transformed data. \textit{J.\ Multivariate Anal.} {\bf 99}, 403--420.

\item Girard, S., Guillou, A.\ and Stupfler, G. (2013). Frontier estimation with kernel regression on high order moments. \textit{J.\ Multivariate Anal.} {\bf  116}, 172--189.

\item Hall, P., Park, B.U.\ and Stern, S.E. (1998). On polynomial estimators of frontiers and boundaries.
  \textit{J.\ Multivariate Anal.} {\bf 66}, 71--98.

\item Hall, P.\ and Van Keilegom, I. (2009). Nonparametric ``regression'' when errors are positioned at end-points. \textit{Bernoulli} {\bf 15}, 614--633.

\item H\"{a}rdle, W., Park, B.U.\ and Tsybakov, A.B. (1995). Estimation of non-sharp support boundaries. \textit{J.\ Multivariate Anal.} {\bf 55}, 205--218.

 \item Jirak, M., Meister, A.\ and Rei{\ss}, M. (2014). Adaptive estimation in nonparametric regression with one-sided errors. \textit{Ann.\ Statist.} {\bf 42}, 1970--2002.

\item Meister, A.\ and  Rei{\ss}, M.  (2013). Asymptotic
equivalence for nonparametric regression with non-regular errors. \textit{Probab.\ Th.\ Rel.\ Fields} {\bf 155}, 201--229.

 \item M\"{u}ller, U.U.\ and Wefelmeyer W. (2010). Estimation in Nonparametric Regression with Non-Regular Errors. \textit{Comm.\ Statist.\ Theory Methods} {\bf 39}, 1619--1629.

\item Neumeyer, N.\ and Van Keilegom, I. (2009). Change-Point Tests for the Error Distribution in Nonparametric Regression. \textit{Scand.\ J.\ Statist.} {\bf 36}, 518--541.
\\
online supporting information available at \\
{http://onlinelibrary.wiley.com/doi/10.1111/j.1467-9469.2009.00639.x/suppinfo}

 \item Picard, D. (1985). Testing and estimating change-points in time series. \textit{Adv. Appl. Probab.} {\bf 17}, 841--867.

 \item
 Rei{\ss}, M.\ and  Selk, L. (2016+). Efficient nonparametric functional estimation for one-sided regression. Bernoulli, to appear.

 \item Resnick, S.I. (2007). \textit{Heavy-Tail Phenomena}. Springer.

\item Simar, L. and Wilson, P.W. (1998). Sensitivity analysis of efficiency scores: how to bootstrap in nonparametric frontier models. {\em Management Science} {\bf 44}, 49--61.

 \item Stephens, M.A. (1976). Asymptotic Results for Goodness-of-Fit Statistics with Unknown Parameters. \textit{Ann.\ Statist.} {\bf 4}, 357--369.

\item van der Vaart, A.W. (2000). \textit{Asymptotic Statistics.}  Cambridge University Press.

 \item van der Vaart, A. W.\ and Wellner, J. A. (1996). \textit{Weak Convergence and Empirical Processes}. Springer, New York.

\item Wilson, P.W. (2003). Testing independence in models of productive efficiency. {\em Journal of Productivity Analysis} {\bf 20}, 361--390.

\end{description}

\end{document}